\documentclass[twocolumn,english]{revtex4-2}
\usepackage[T1]{fontenc}
\usepackage[latin9]{inputenc}
\usepackage[active]{srcltx}
\usepackage{babel}
\usepackage{amstext}
\usepackage{amssymb}
\usepackage{graphicx}
\usepackage{wasysym}
\usepackage[unicode=true,pdfusetitle,
 bookmarks=true,bookmarksnumbered=false,bookmarksopen=false,
 breaklinks=false,pdfborder={0 0 1},backref=false,colorlinks=false]
 {hyperref}

\makeatletter

\providecommand{\tabularnewline}{\\}

\makeatother

\begin{document}
\title{A nonequilibrium system on a restricted scale-free network}
\author{R. A. Dumer }
\email{rafaeldumer@fisica.ufmt.br}

\affiliation{Instituto de Física - Universidade Federal de Mato Grosso, 78060-900,
Cuiabá, Mato Grosso, Brazil.}
\author{M. Godoy}
\email{mgodoy@fisica.ufmt.br}

\affiliation{Instituto de Física - Universidade Federal de Mato Grosso, 78060-900,
Cuiabá, Mato Grosso, Brazil.}
\begin{abstract}
The nonequilibrium Ising model on a restricted scale-free network
has been studied with one- and two-spin flip competing dynamics employing
Monte Carlo simulations. The dynamics present in the system can be
defined by the probability $q$ in which the one-spin flip process
simulate the contact with a heat bath at a given temperature $T$,
and with a probability $(1-q)$ the two-spin flip process mimics the
system subjected to an external flux of energy into it. The system
network is described by a power-law degree distribution in the form
$P(k)\sim k^{-\alpha}$, and the restriction is made by fixing the
maximum, $k_{m}$, and minimum, $k_{0}$, degree on distribution for
the whole network size. This restriction keeps finite the second and
fourth moment of degree distribution, allowing us to obtain a finite
critical point for any value of $\alpha$. For these critical points,
we have calculated the thermodynamic quantities of the system, such
as, the total $\textrm{\ensuremath{\textrm{m}_{\textrm{N}}^{\textrm{F}}}}$
and staggered $\textrm{\ensuremath{\textrm{m}_{\textrm{N}}^{\textrm{AF}}}}$
magnetizations per spin, susceptibility $\textrm{\ensuremath{\chi_{\textrm{N}}}}$,
and reduced fourth-order Binder cumulant $\textrm{\ensuremath{\textrm{U}_{\textrm{N}}}}$,
for several values of lattice size $N$ and exponent $1\le\alpha\le5$.
Therefore, the phase diagram was built and a self-organization phenomena
is observed from the transitions between antiferromagnetic $AF$ to
paramagnetic $P$, and $P$ to ferromagnetic $F$ phases. Using the
finite-size scaling theory, we also obtained the critical exponents
for the system, and a mean-field critical behavior is observed, exhibiting
the same universality class of the system on the equilibrium and out
of it.
\end{abstract}
\maketitle

\section{Introduction}

The dynamic evolution of equilibrium systems is related to the fact
that the transition rates of its states obey the principle of microscopic
reversibility. Otherwise, without the advanced tooling as proposed
by Gibbs in the equilibrium scene \citep{1}, nonequilibrium systems
have aroused the interest of researchers in finding out phase transitions
with the particularities of continuous phase transitions of reversible
systems. One kind of the nonequilibrium system is those subjected
to two dynamics in competition \citep{2,3}. These systems are described
by a master equation that involves the sum of the operators on each
present process and generally each of these processes separately obeys
the principle of microscopic reversibility. However, the combination
of these processes may not satisfy the detailed balance and the system
will be forced out of equilibrium.

In the last decades, the computerization of data acquisition on large
networks, make raised the possibility of understanding the dynamical
and topological stability of its networks. From that databases, the
result is that large networks that span fields as diverse as the World
Wide Web (WWW) or actors that have acted in a movie together, self-organize
into a scale-free state \citep{4,5}. This means that independent
of the system and its constituents, the probability $P(k)$ that a
vertex interacts with $k$ other vertices in the network, decay as
a power law, i.e., $P(k)\sim k^{-\alpha}$. Barabási and Albert \citep{5}
incorporating growth and preferential attachment on its network model,
were able to obtain this scale invariance, not present in the previous
random \citep{6} and small-world networks \citep{7}. These models
and their interesting ability to describe real networks instigated
the curiosity of researchers to know what would be the behavior of
physical systems in complex networks \citep{8,9,10,11,12}. Among
these, we can highlight the simple but powerful Ising model, comprising
both exact \citep{13,14} and computational \citep{15,16,17,18} or
approximate \citep{19,20,21} results for the critical behavior on
arbitrary networks.

In the same way, the study of nonequilibrium physical systems has
been spreading and continuous phase transitions, characteristic of
equilibrium systems is observed \citep{22,23}. Moreover, the same
critical exponents have been obtained in reversible and irreversible
systems, that is, they belong to the same universality class, acting
as proof of what was conjectured by Grinstein \emph{et al}. \citep{24},
in which says that any nonequilibrium\emph{ }stochastic spin system
with spin-flip dynamics and up-down symmetry belongs to the same universality
class. The Ising model with complex networks is already being studied
with competing dynamics, analytically in 1D \citep{25}, by Monte
Carlo simulations in 2D \citep{26}, and by Gaussian model in 3D \citep{27}.
However, these studies were made only for small-world networks, and
by Monte Carlo simulations a mean-field critical behavior is obtained,
characteristic of equilibrium systems with random interactions and
convergent fourth moment of its network degree distribution \citep{13,14,15,17}.
Another interesting feature of that nonequilibrium systems is the
self-organization phenomena between antiferromagnetic $AF$ to paramagnetic
$P$, and $P$ to ferromagnetic $F$ phase transitions, as a function
of competition parameter \citep{2,3,22,23}.

With this in mind, in the present work, we have investigated the Ising
model on a restricted scale-free network, where each site of the network
is occupied by a spin variable that can assume values $\pm1$. Divided
into two sublattices, the connections between them in the network
are made by the site interactions, and the degree distribution of
the network obey a power-law distribution, with fixed values of minimum
and maximum degree. The system is in a nonequilibrium regime by competing
between two reagent dynamic processes that do not conserve the order
parameter: with competition probability $q$, the one-spin flip process
simulates the system in contact with a heat bath at temperature $T$,
and with probability $1-q$, the two-spin flip process mimics the
system subjected to an external flux into it. Thus, here we have investigated
the phase transitions of the system and verified if the phase diagrams
present the same topology of systems with these same dynamics \citep{23,26},
and in addition, the critical exponents carrying the universality
class of the system, is compared with previous works at equilibrium
system \citep{16}.

This article is organized as follows: In Section \ref{sec:Model},
we describe the network used and the Hamiltonian model of the system.
In Section \ref{sec:Monte-Carlo-simulations}, we present the Monte
Carlo simulation method, some details concerning the simulation procedures,
and the thermodynamic quantities of the system, also necessary for
the application of FSS analysis. The behavior of thermodynamic quantities,
phase diagrams, and critical exponents are described in Section \ref{sec:Results}.
Finally, in Section \ref{sec:Conclusions}, we present our conclusions.

\section{Model\label{sec:Model}}

\begin{figure}
\begin{centering}
\includegraphics[scale=0.7]{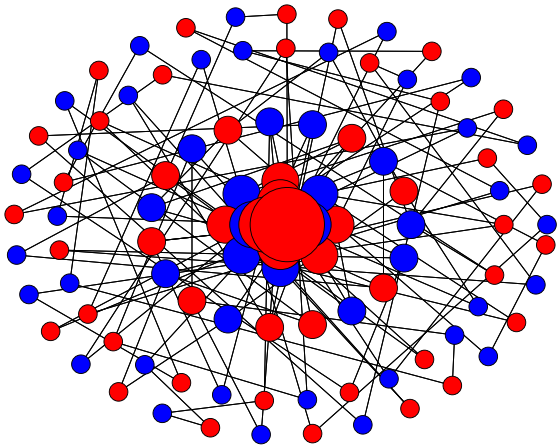}
\par\end{centering}
\caption{{\footnotesize{}Schematic representation of the restricted scale-free
network. Red circles indicate the sites on one of the sublattices,
blue circles are the sites on the other sublattice, and the black
solid lines are the connections between the two sublattices. The size
of the circles is proportional to the degree of sites, varying from
$k_{0}=2$ to $k_{m}=8$ in the distribution with $\alpha=3$, and
$N=10^{2}$. \label{fig:1}}}
\end{figure}

The Ising model studied in this work has $N$ spins $\sigma_{i}=\pm1$
on a restricted scale-free network and ferromagnetic interaction of
strength $J_{ij}$. The degree distribution on the network follows
the power-law $P(k)\sim k^{-\alpha}$ and to distribute the connections
between the sites, we have used the same procedures shown in the paper
\citep{16}. In order to construct a scale-free network with always
convergent second and fourth moments on its degree distribution and
arbitrary value of $\alpha$. For that, we first define minimum $k_{0}$
and maximum $k_{m}$ degree, and the exponent $\alpha$ of the distribution.
The next procedure is to calculate the normalization constant of the
distribution, $A=\sum_{k=k_{0}}^{k_{m}}k^{\alpha}$, and found the
smaller network size that we can use and guarantee the degree distribution,
$N_{0}=k_{m}^{\alpha}/A$. With these values, we create a set of site
numbers, $\{N_{k}\}$, and that will have the respective degrees $k$,
where $N_{k}=AN/k^{\alpha}$. On that distribution of connections,
we have divided the network into two sublattices, where one sublattice
plays the role of central spins, while the other sublattice contains
the spins in which the central spins can connect. Thus, starting with
the lowest degree $k_{0}$, connections of each $N_{k_{0}}$ sites
are randomly created connecting the two sublattices, and it was made
until reach degree $k_{m}$ and the whole set $\{N_{k}\}$ will be
visited. An example of that construction can be seen in Fig. \ref{fig:1}
which was chosen $\alpha=3$, $k_{0}=2$, $k_{m}=8$ and $N=10^{2}$.
In Fig. \ref{fig:1}, the sites in the middle of the figure are the
more connected, while the peripheral sites are the less connected,
and sites from the blue sublattice are only connected with sites from
the red sublattice.

Based on this construction, in the course of this work, we have selected
the integer values of $1\le\alpha\le5$, $k_{0}=4$, $k_{m}=10$,
and network size $(32)^{2}\le N\le(256)^{2}$ to study the nonequilibrium
Ising model. The ferromagnetic Ising spin energy is described by the
Hamiltonian on the form

\begin{equation}
\mathcal{H}=-\sum_{\left\langle i,j\right\rangle }J_{ij}\sigma_{i}\sigma_{j}\label{eq:1}
\end{equation}
where the sum is over all pair of spins, and $J_{ij}$ is the ferromagnetic
interaction, assuming the value of unity if sites $i$ and $j$ interact
between the sublattices.

In the nonequilibrium system presented here, let $p(\{\sigma\},t)$
be the probability of finding the system in the state $\{\sigma\}=\{\sigma_{1},...,\sigma_{i},...,\sigma_{j},...\sigma_{N}\}$
at time $t$, the motion equation for the probability of states evolves
in time according to the master equation 

\begin{equation}
\frac{d}{dt}p(\{\sigma\},t)=qG+(1-q)D,\label{eq:2}
\end{equation}
where $qG$ represents the one-spin flip process, relaxing the spins
in contact with a heat bath at temperature $T$, favoring the lowest
energy state of the system, and has probability $q$ to occur. On
the other hand, the $(1-q)D$ denotes the two-spin flip process, in
which the energy of the system increases by one external flow of energy
into it, and has a probability $(1-q)$ to occur. $G$ and $D$ are
described as follows:

\begin{equation}
\begin{array}{ccc}
G= & \sum_{i,\{\sigma'\}}\left[W(\sigma_{i}\to\sigma_{i}')p(\{\sigma\},t)+\right.\\
 & \left.-W(\sigma_{i}'\to\sigma_{i})p(\{\sigma'\},t)\right] & ,
\end{array}\label{eq:3}
\end{equation}

\begin{equation}
\begin{array}{ccc}
D= & \sum_{i,j,\{\sigma'\}}\left[W(\sigma_{i}\sigma_{j}\to\sigma_{i}'\sigma_{j}')p(\{\sigma\},t)+\right.\\
 & \left.-W(\sigma_{i}'\sigma_{j}'\to\sigma_{i}\sigma_{j})p(\{\sigma'\},t)\right] & ,
\end{array}\label{eq:4}
\end{equation}
where $\{\sigma'\}$ is the spin configuration after spin flipping,
$W(\sigma_{i}\to\sigma_{i}')$ is the transition rate between the
states in the one-spin flip process, and $W(\sigma_{i}\sigma_{j}\to\sigma_{i}'\sigma_{j}')$
the transition rate between the states in the two-spin flip process.

\section{Monte Carlo simulations\label{sec:Monte-Carlo-simulations}}

\begin{figure}
\begin{centering}
\includegraphics[scale=0.33]{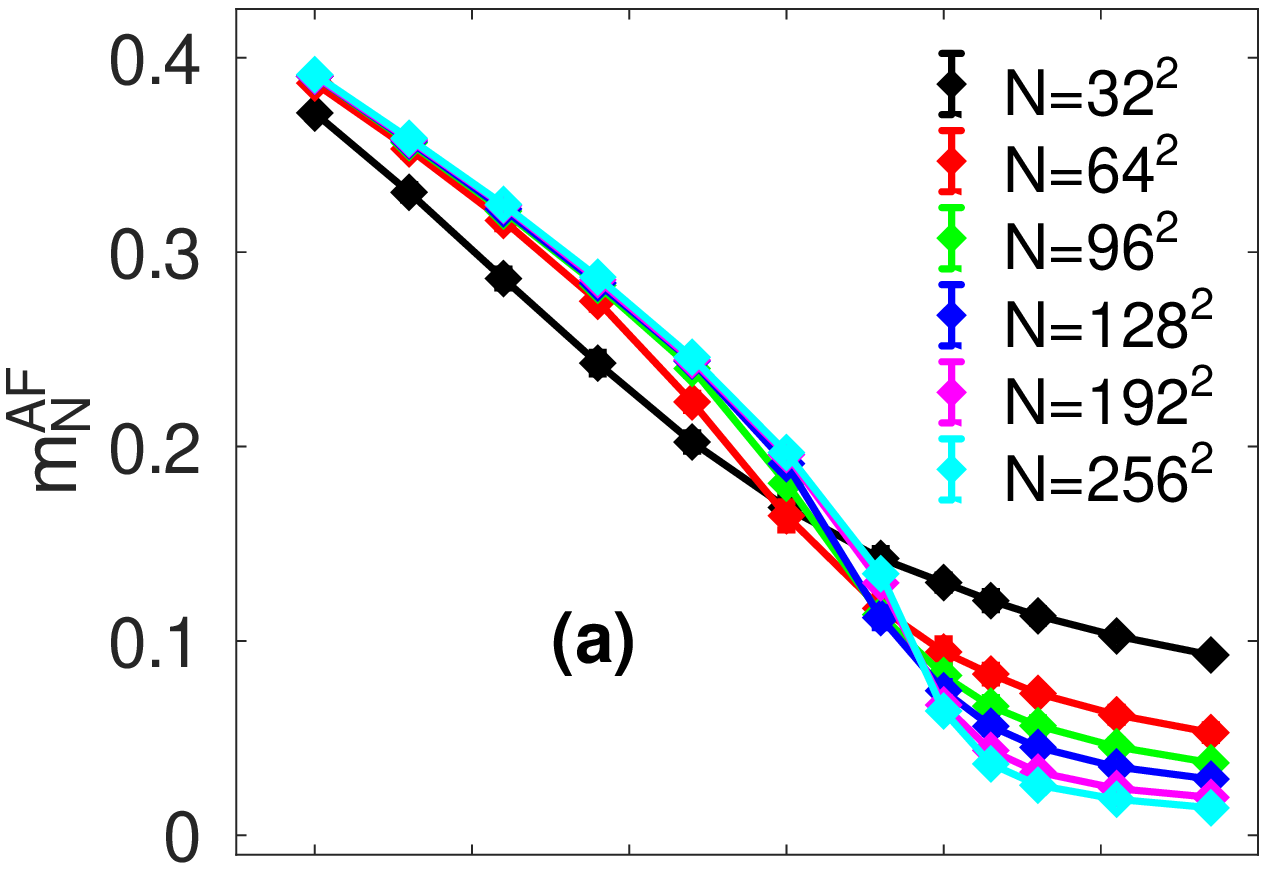}\includegraphics[scale=0.33]{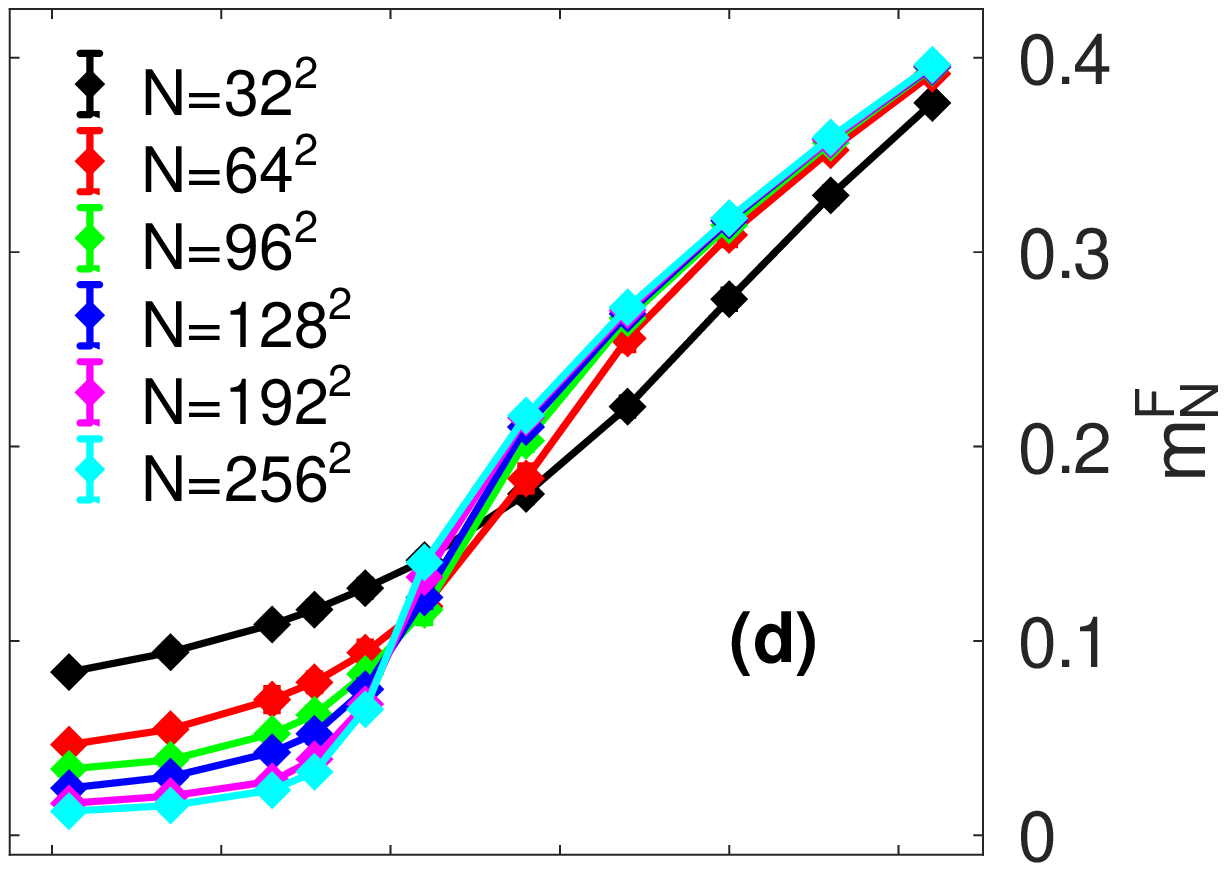}
\par\end{centering}
\begin{centering}
\includegraphics[scale=0.33]{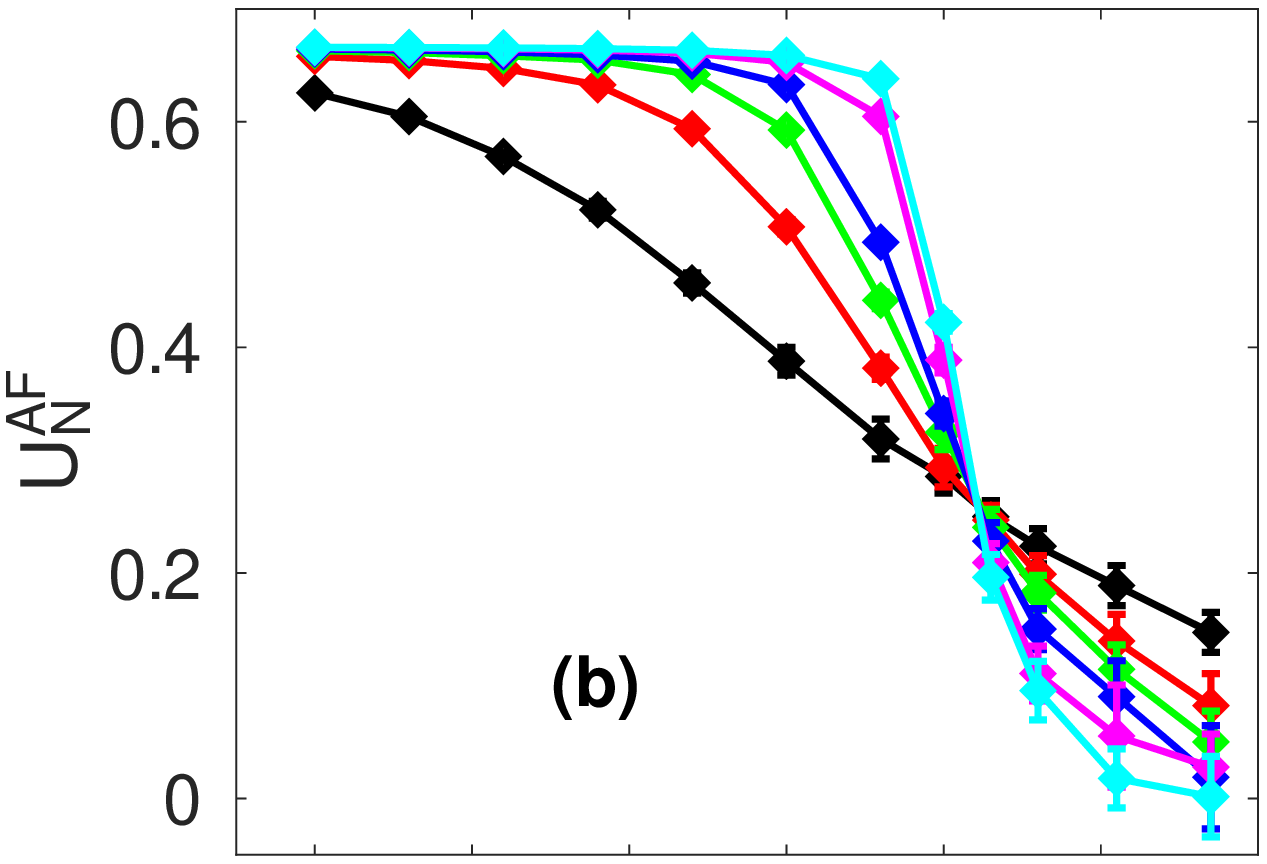}\includegraphics[scale=0.33]{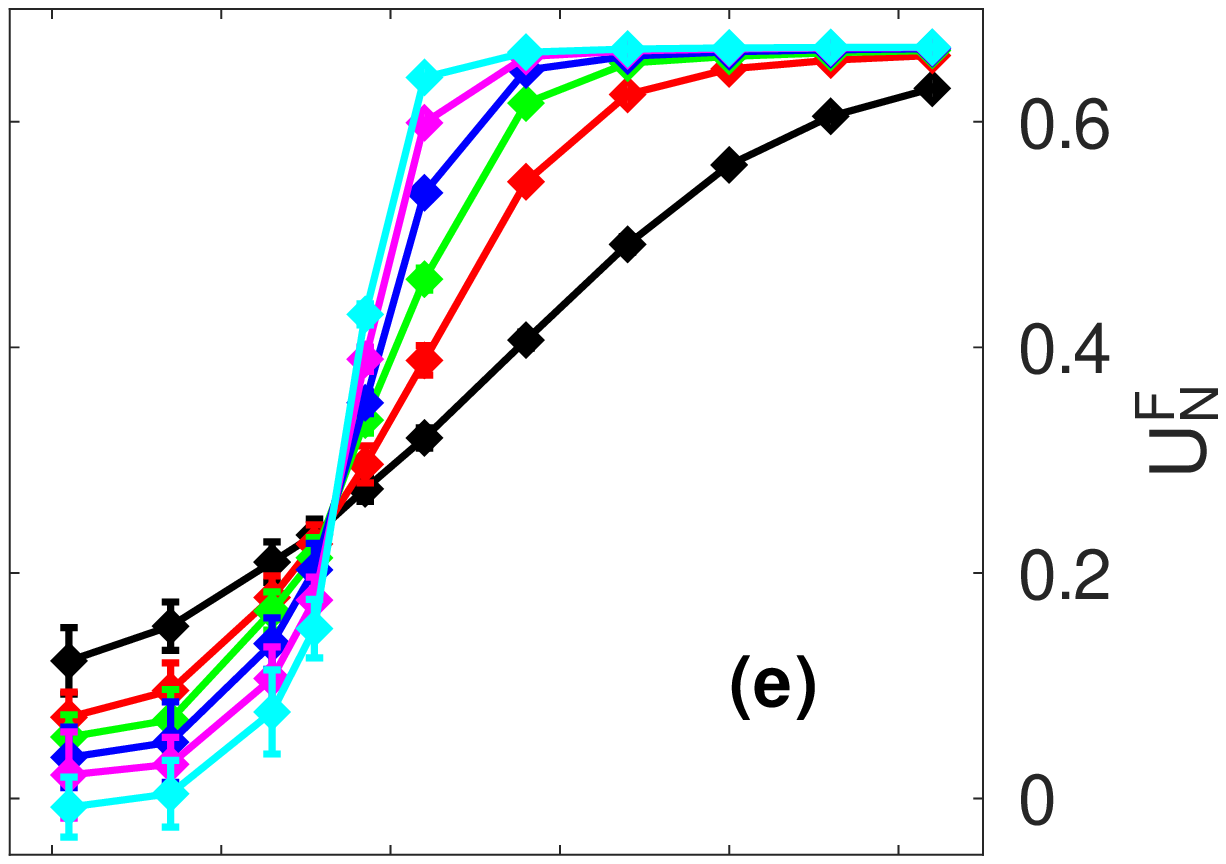}
\par\end{centering}
\begin{centering}
\includegraphics[scale=0.33]{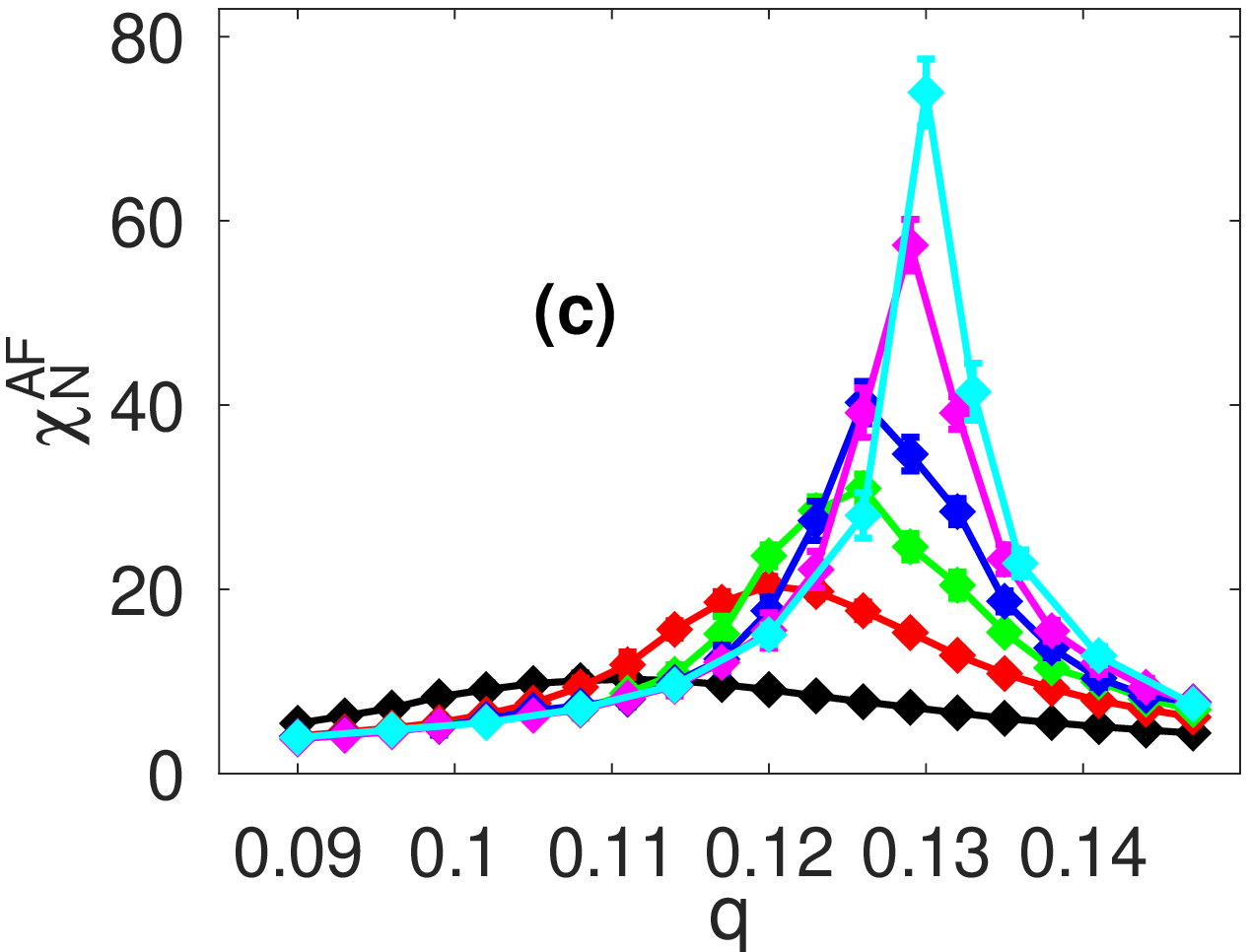}\includegraphics[scale=0.33]{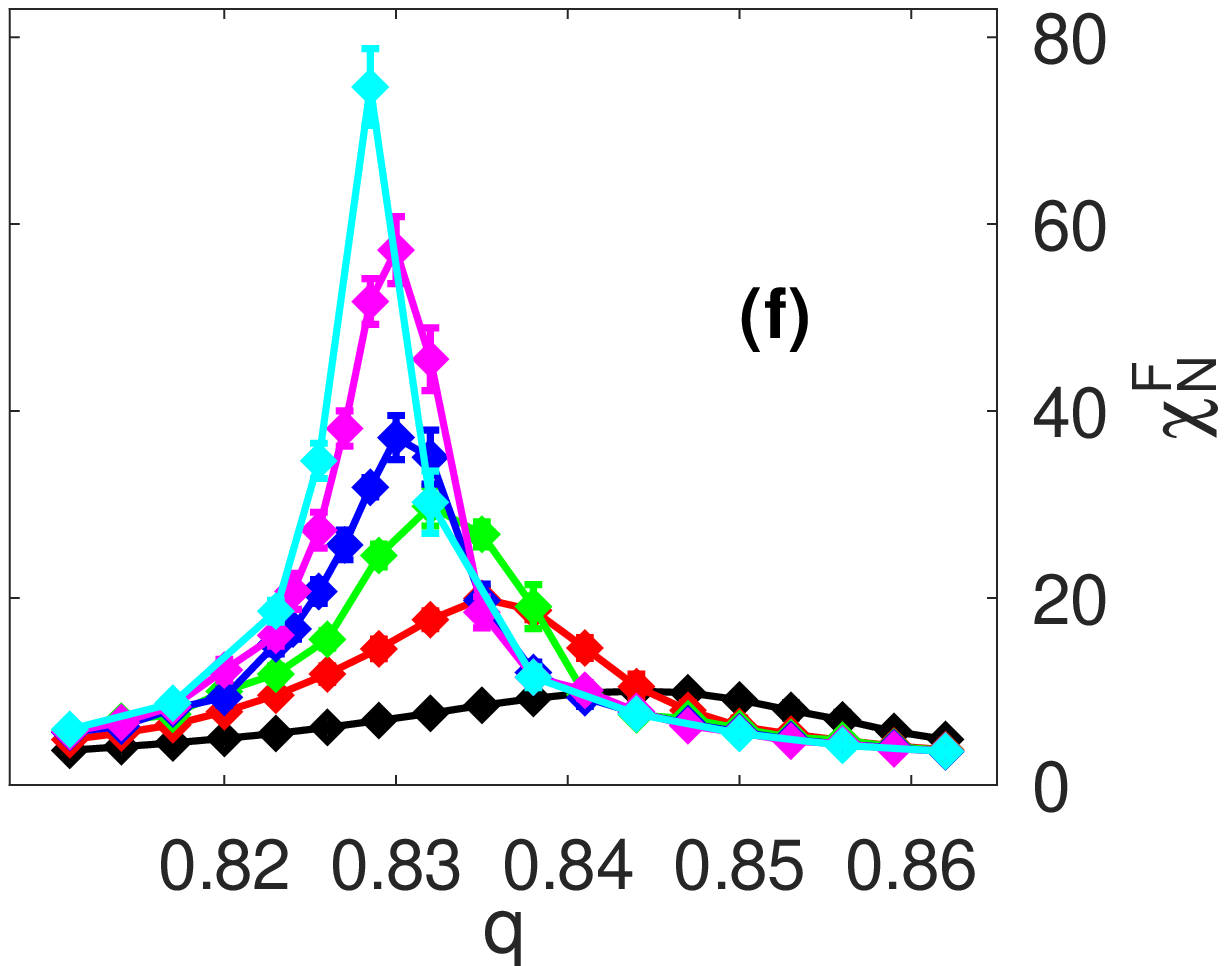}
\par\end{centering}
\caption{{\footnotesize{}Behavior of the thermodynamic quantities $\textrm{m}_{\textrm{L}}$,
$\textrm{U}_{\textrm{L}}$, and $\textrm{\ensuremath{\chi}}_{\textrm{L}}$as
a function of $q$ for different network sizes as presented in the
figures. In the $AF-P$ transition: (a) $\textrm{m}_{\textrm{L}}^{\textrm{AF}}$,
(b) $\textrm{U}_{\textrm{L}}^{\textrm{AF}}$, and (c) $\textrm{\ensuremath{\chi}}_{\textrm{L}}^{\textrm{AF}}$curves.
In the $F-P$ transition: (d) $\textrm{m}_{\textrm{L}}^{\textrm{F}}$
, (e) $\textrm{U}_{\textrm{L}}^{\textrm{F}}$, and (f) $\textrm{\ensuremath{\chi}}_{\textrm{L}}^{\textrm{F}}$curves.
Here, we have fixed values of $\alpha=1$, $k_{0}=4$, $k_{m}=10$,
$T=1$.\label{fig:2}}}
\end{figure}

In the simulation of the system specified by the Hamiltonian in Eq.
(\ref{eq:1}), we always have chosen the initial state of the system
with all spin states at random, and a new configuration is generated
by the following Markov process: for a given temperature $T$, competition
probability $q$, distribution exponent $\alpha$, network size $N$,
and minimum $k_{0}$ and maximum $k_{m}$ degree, we choose at random
a spin $\sigma_{i}$ in network, and generate a random number $\xi$
between zero and one. If $\xi\le q$, we choose the one-spin flip
process, in which the flipping probability is dependent of $W(\sigma_{i}\to\sigma_{i}^{\prime})$
and given by the Metropolis prescription:

\begin{equation}
W(\sigma_{i}\to\sigma_{i}^{\prime})=\left\{ \begin{array}{cccc}
e^{\left(-\Delta E_{i}/k_{B}T\right)} & \textrm{if} & \Delta E_{i}>0\\
1 & \textrm{if} & \Delta E_{i}\le0 & ,
\end{array}\right.\label{eq:5}
\end{equation}
where $\Delta E_{i}$ is the change in energy after flipping the spin,
$\sigma_{i}\to\sigma_{i}^{\prime}$, $k_{B}$ is the Boltzmann constant,
and $T$ the temperature of the system. Thus, the acceptance of a
new state is guaranteed if $\Delta E_{i}\le0$, but, in the case where
$\Delta E>0$ the acceptance is pondered by the probability $\exp\left(-\Delta E_{i}/k_{B}T\right)$
and just guaranteed if by choosing a random number, $0<\xi_{1}<1$,
it is $\xi_{1}\le\exp\left(-\Delta E_{i}/k_{B}T\right)$. On the other
hand, if none of these conditions are satisfied, we do not change
the state of the system. Now, if $\xi>q$ the two-spin flip process
is chosen, and in addition to the spin $\sigma_{i}$ we also randomly
choose one of its neighbors $\sigma_{j}$, and these two spins are
flipping simultaneously according to transition rate $W(\sigma_{i}\sigma_{j}\to\sigma_{i}'\sigma_{j}')$
given by

\begin{equation}
W(\sigma_{i}\sigma_{j}\to\sigma_{i}'\sigma_{j}')=\left\{ \begin{array}{c}
0\\
1
\end{array}\begin{array}{c}
\textrm{if}\\
\textrm{if}
\end{array}\begin{array}{cc}
\Delta E_{ij}\le0\\
\Delta E_{ij}>0 & ,
\end{array}\right.\label{eq:6}
\end{equation}
where $\Delta E_{ij}$ is the change in the energy after flipping
the spins $\sigma_{i}$ and $\sigma_{j}$, and consequently, in this
process, the new state is only accepted if $\Delta E_{ij}>0$.

Repeating the Markov process $N$ times, we have one Monte Carlo Step
(MCS). In our simulations, we have waited for $10^{4}$ MCS to the
system reach the stationary state, in the whole the network sizes
and adjustable parameters. In order to calculate the thermal averages
of the interest quantities, we used more $4\times10^{4}$ MCS, and
the average over samples was done using $10$ independent samples
for any configuration. 

The measured thermodynamic quantities in our simulations are: magnetization
per spin $\textrm{m}_{\textrm{N}}^{\textrm{F}}$, staggered magnetization
per spin $\textrm{m}_{\textrm{N}}^{\textrm{AF}}$, magnetic susceptibility
$\textrm{\ensuremath{\chi}}_{\textrm{N}}$ and reduced fourth-order
Binder cumulant $\textrm{U}_{\textrm{N}}$:

\begin{equation}
\textrm{m}_{\textrm{N}}^{\textrm{F}}=\frac{1}{N}\left[\left\langle \sum_{i=1}^{N}\sigma_{i}\right\rangle \right],\label{eq:7}
\end{equation}

\begin{equation}
\textrm{m}_{\textrm{N}}^{\textrm{AF}}=\frac{1}{N}\left[\left\langle \sum_{i=1}^{N}(-1)^{(r+c)}\sigma_{i}\right\rangle \right],\label{eq:8}
\end{equation}

\begin{equation}
\textrm{\ensuremath{\chi}}_{\textrm{N}}=\frac{N}{k_{B}T}\left[\left\langle m^{2}\right\rangle -\left\langle m\right\rangle ^{2}\right],\label{eq:9}
\end{equation}

\begin{equation}
\textrm{U}_{\textrm{N}}=1-\frac{\left[\left\langle m^{4}\right\rangle \right]}{3\left[\left\langle m^{2}\right\rangle ^{2}\right]},\label{eq:10}
\end{equation}
where $\left[\ldots\right]$ representing the average over the samples,
and $\left\langle \ldots\right\rangle $ the thermal average over
the MCS in the stationary state. To facilitate the calculation of
$\textrm{m}_{\textrm{N}}^{\textrm{AF}}$, the sites on the network
are labeled as if we had a square lattice, $N=L^{2}$, in this way,
$r$ and $c$ are the row and column of the site $i$, respectively.
In Eqs. (\ref{eq:9}) and (\ref{eq:10}), $m$ can be used to represent
$\textrm{\ensuremath{\textrm{m}_{\textrm{N}}^{\textrm{F}}}}$ or $\textrm{\ensuremath{\textrm{m}_{\textrm{N}}^{\textrm{AF}}}}$.

In the vicinity of the stationary critical point $\lambda_{c}$, the
Eqs. (\ref{eq:7}), (\ref{eq:8}), (\ref{eq:9}) and (\ref{eq:10})
obey the following finite-size scaling relations \citep{28}:

\begin{equation}
\textrm{m}_{\textrm{N}}=N^{-\beta/\nu}m_{0}(N^{1/\nu}\epsilon),\label{eq:11}
\end{equation}

\begin{equation}
\textrm{\ensuremath{\chi}}_{\textrm{N}}=N^{\gamma/\nu}\chi_{0}(N^{1/\nu}\epsilon),\label{eq:12}
\end{equation}

\begin{equation}
\textrm{U}_{\textrm{N}}^{\prime}=N^{1/\nu}\frac{U_{0}^{\prime}(N^{1/\nu}\epsilon)}{\lambda_{c}},\label{eq:13}
\end{equation}
where $\epsilon=(\lambda-\lambda_{c})/\lambda_{c}$ ($\lambda$ and
$\lambda_{c}$ can be used $T$ or $q$), and $\beta$, $\gamma$
and $\nu$ are the critical exponents related the magnetization, susceptibility
and length correlation, respectively. The functions $m_{0}(N^{1/\nu}\epsilon)$,
$\chi_{0}(N^{1/\nu}\epsilon)$ and $U_{0}(N^{1/\nu}\epsilon)$ are
the scaling functions.

Using the data from simulations for the network sizes $(32)^{2}\le N\le(256)^{2}$
in the Eqs. (\ref{eq:11}), (\ref{eq:12}) and (\ref{eq:13}), we
have obtained the critical exponents ratio $\beta/\nu$, $\gamma/\nu$,
and $\nu^{-1}$ from the slope of the straight lines in the log-log
plot of $\textrm{m}_{\textrm{N}}(\lambda_{c})$, $\textrm{\ensuremath{\chi}}_{\textrm{N}}(\lambda_{c})$,
and $\textrm{U}_{\textrm{N}}^{\prime}(\lambda_{c})$ (derivative of
$\textrm{U}_{\textrm{N}}$) as a function of $N$. Besides that, we
also used data collapse from scaling functions to estimate the critical
exponent values.

\section{Results\label{sec:Results}}

\begin{figure}
\begin{centering}
\includegraphics[bb=10bp 26bp 385bp 300bp,clip,scale=0.33]{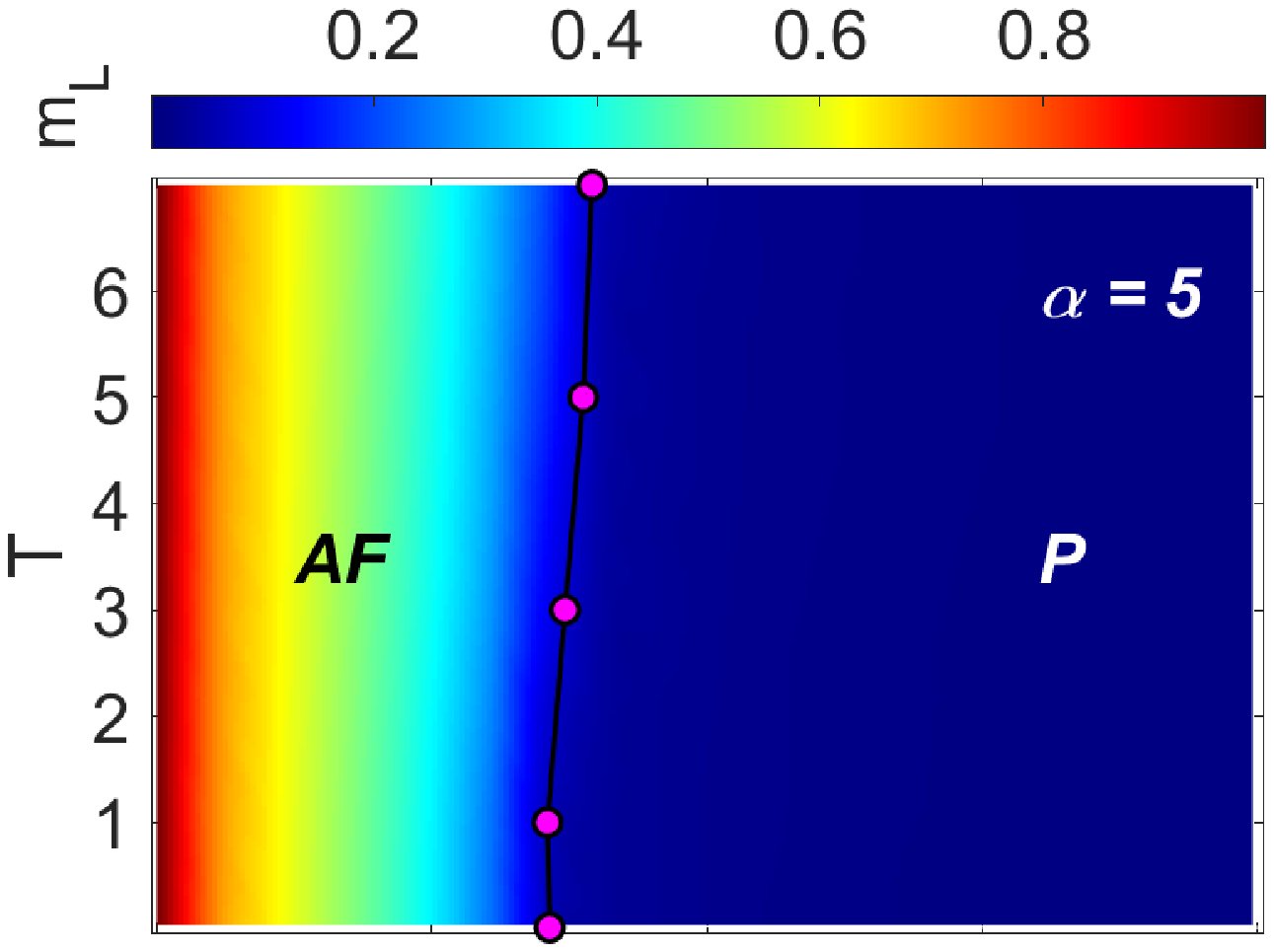}\includegraphics[bb=53bp 26bp 385bp 300bp,clip,scale=0.33]{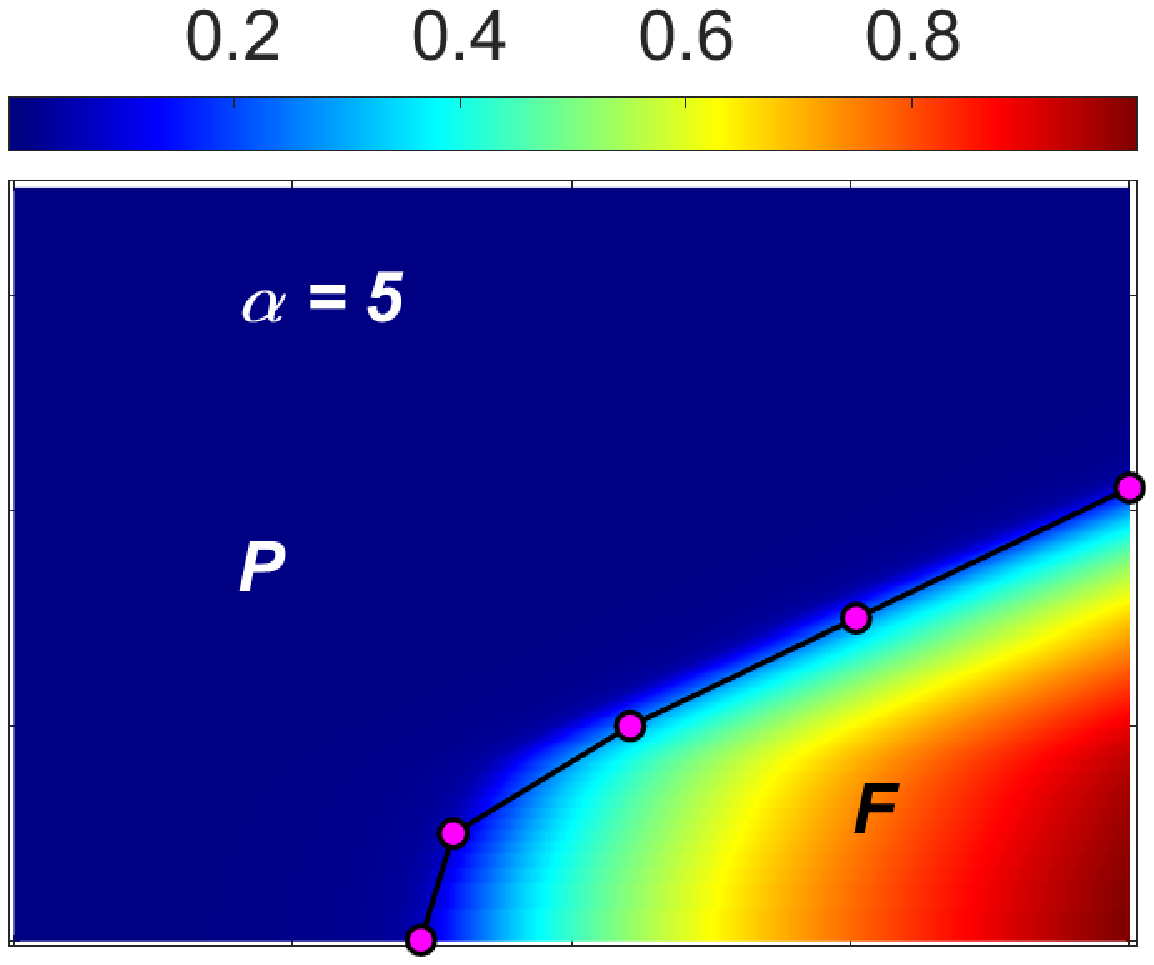}
\par\end{centering}
\begin{centering}
\includegraphics[bb=10bp 30bp 385bp 292bp,clip,scale=0.33]{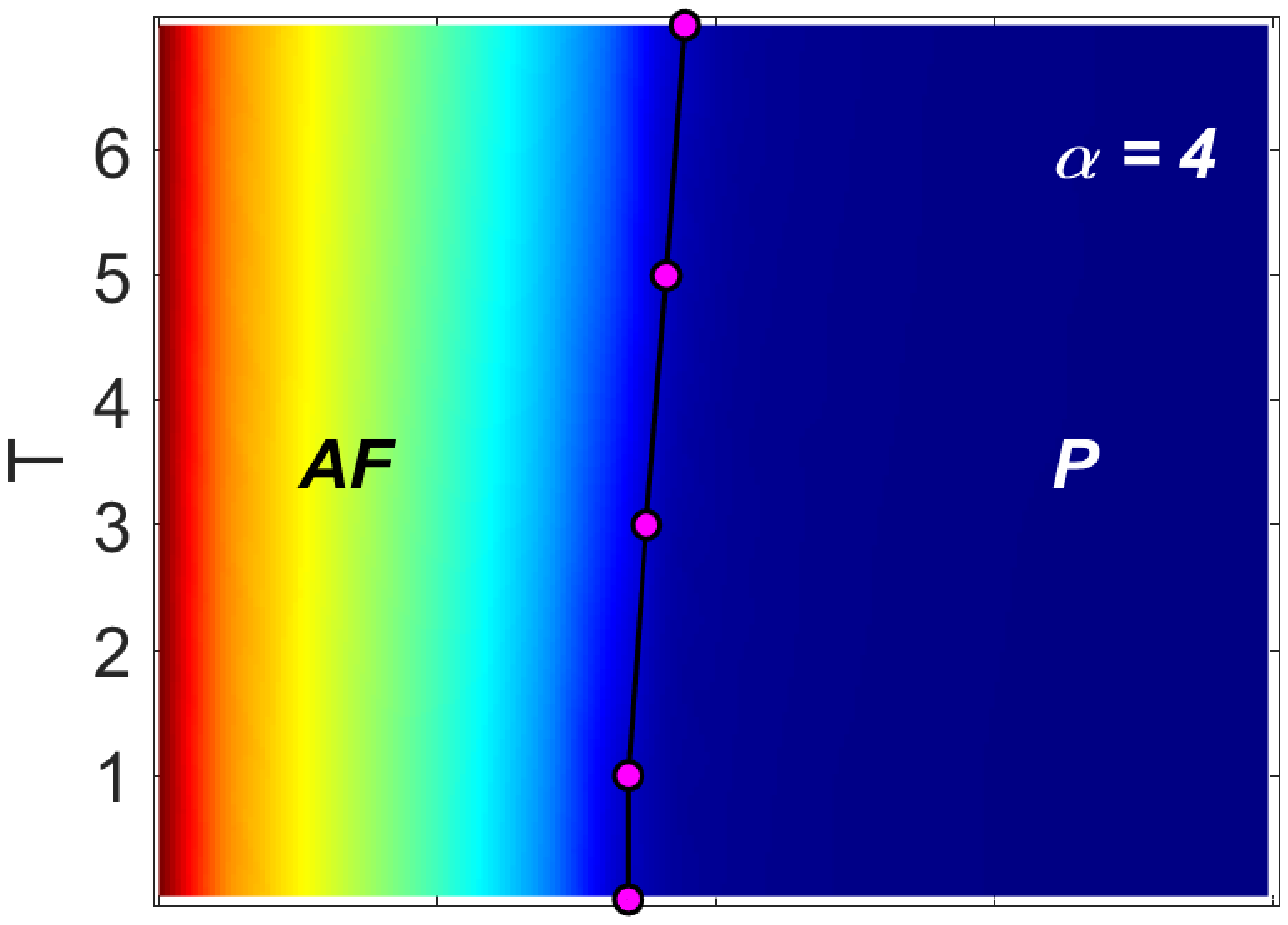}\includegraphics[bb=53bp 30bp 385bp 292bp,clip,scale=0.33]{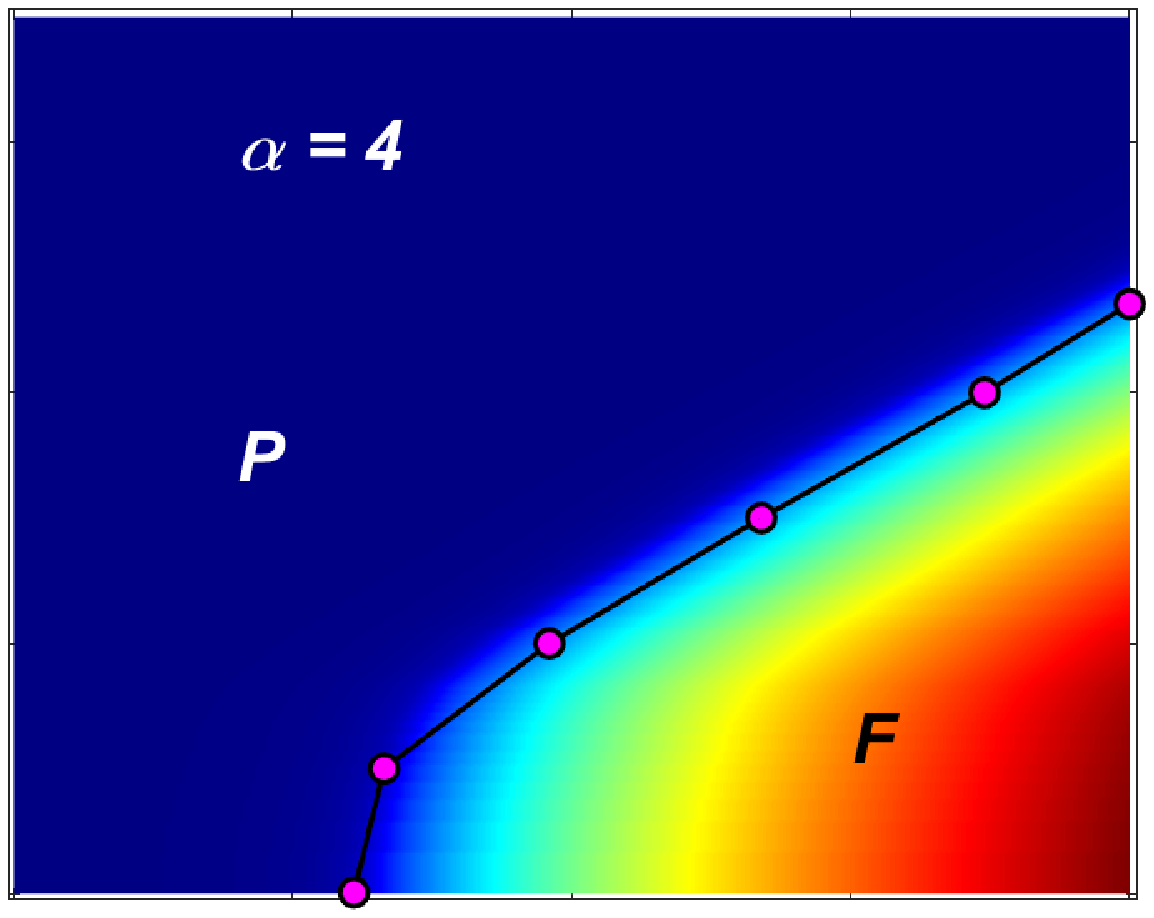}
\par\end{centering}
\begin{centering}
\includegraphics[bb=10bp 30bp 385bp 292bp,clip,scale=0.33]{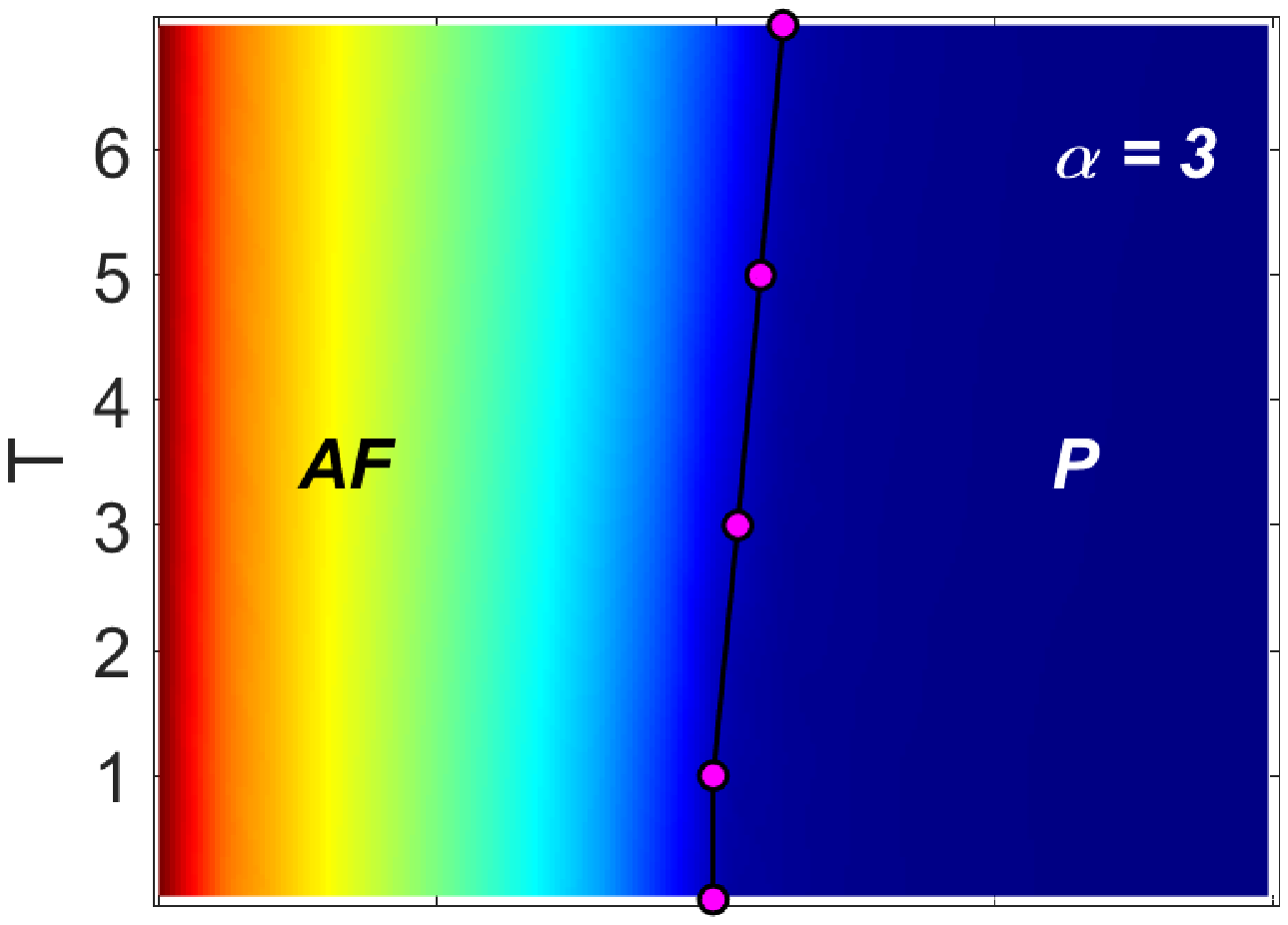}\includegraphics[bb=53bp 30bp 385bp 292bp,clip,scale=0.33]{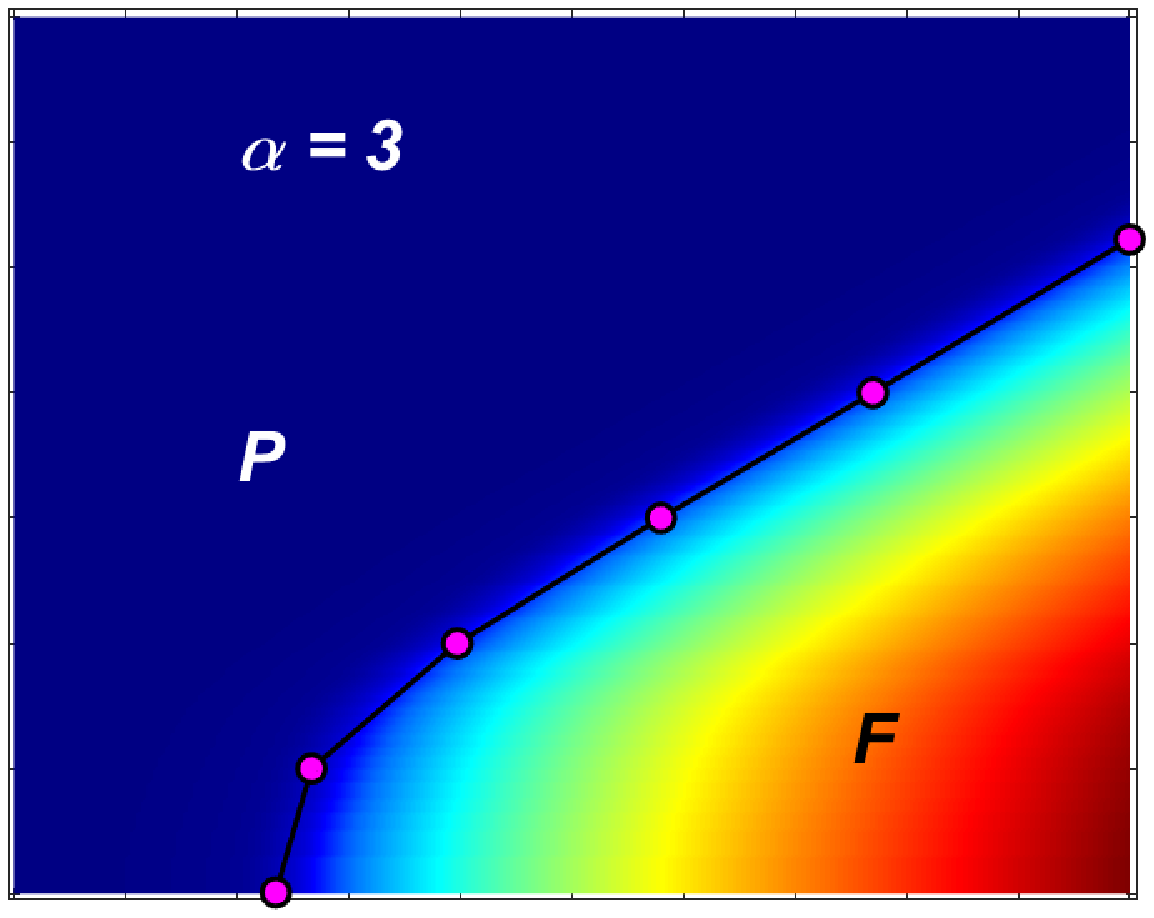}
\par\end{centering}
\begin{centering}
\includegraphics[bb=10bp 30bp 385bp 292bp,clip,scale=0.33]{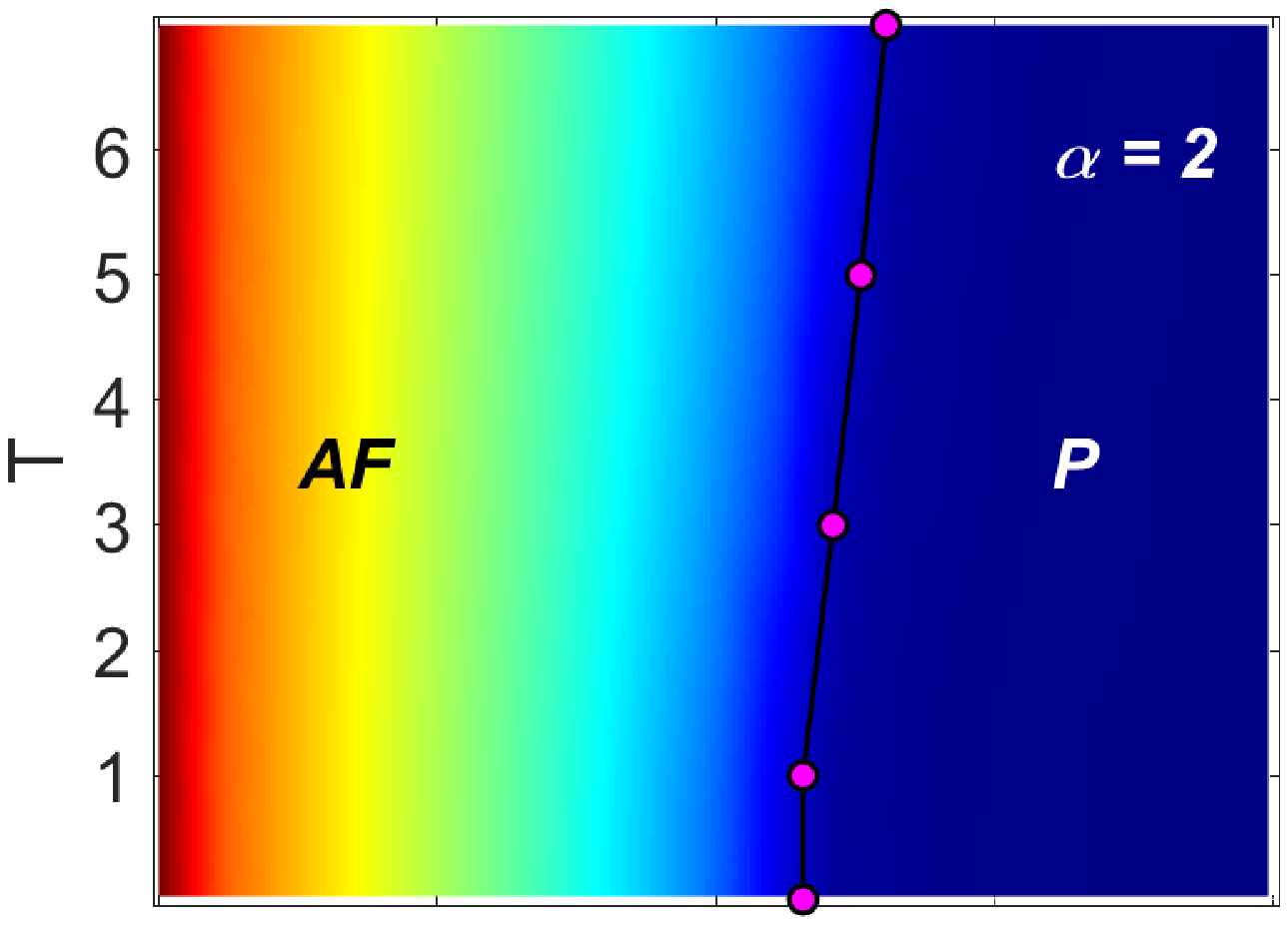}\includegraphics[bb=53bp 30bp 385bp 292bp,clip,scale=0.33]{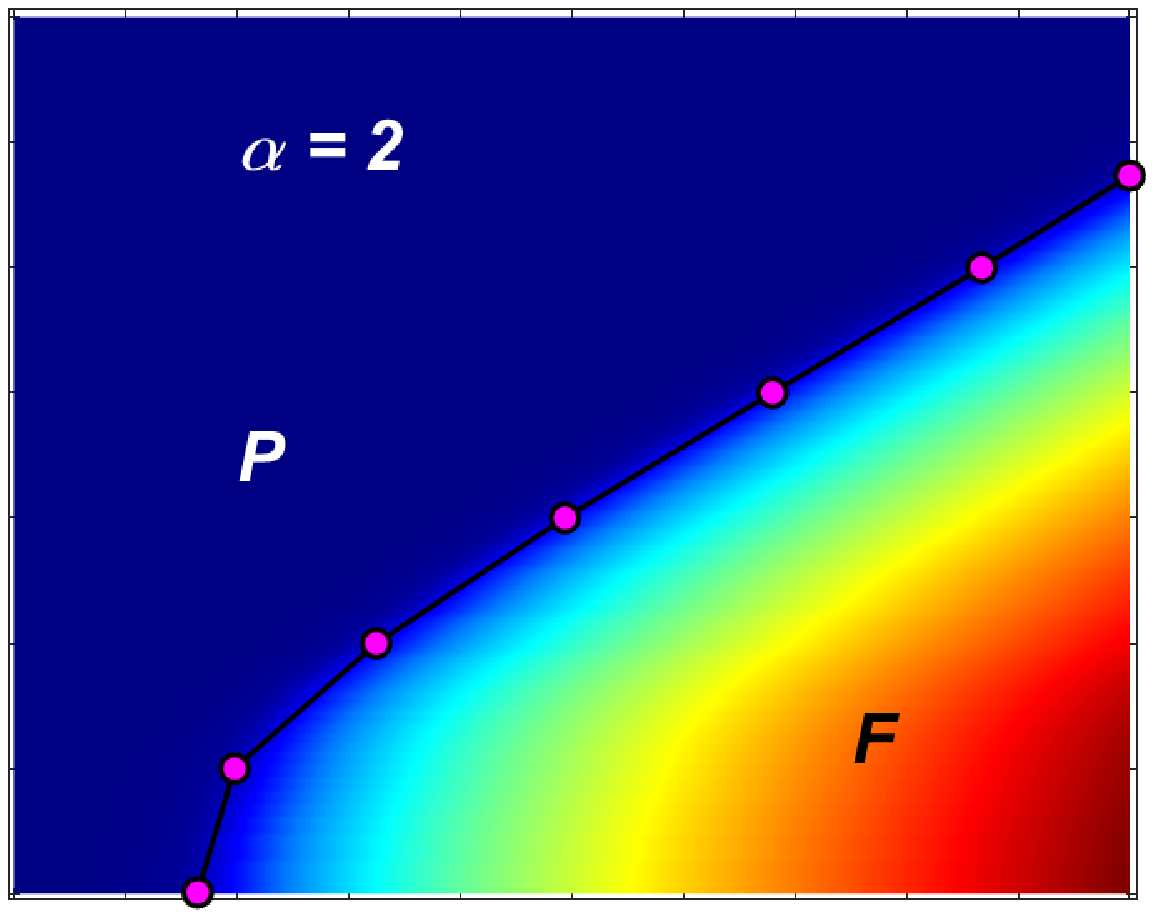}
\par\end{centering}
\begin{centering}
\includegraphics[bb=10bp 0bp 385bp 292bp,clip,scale=0.33]{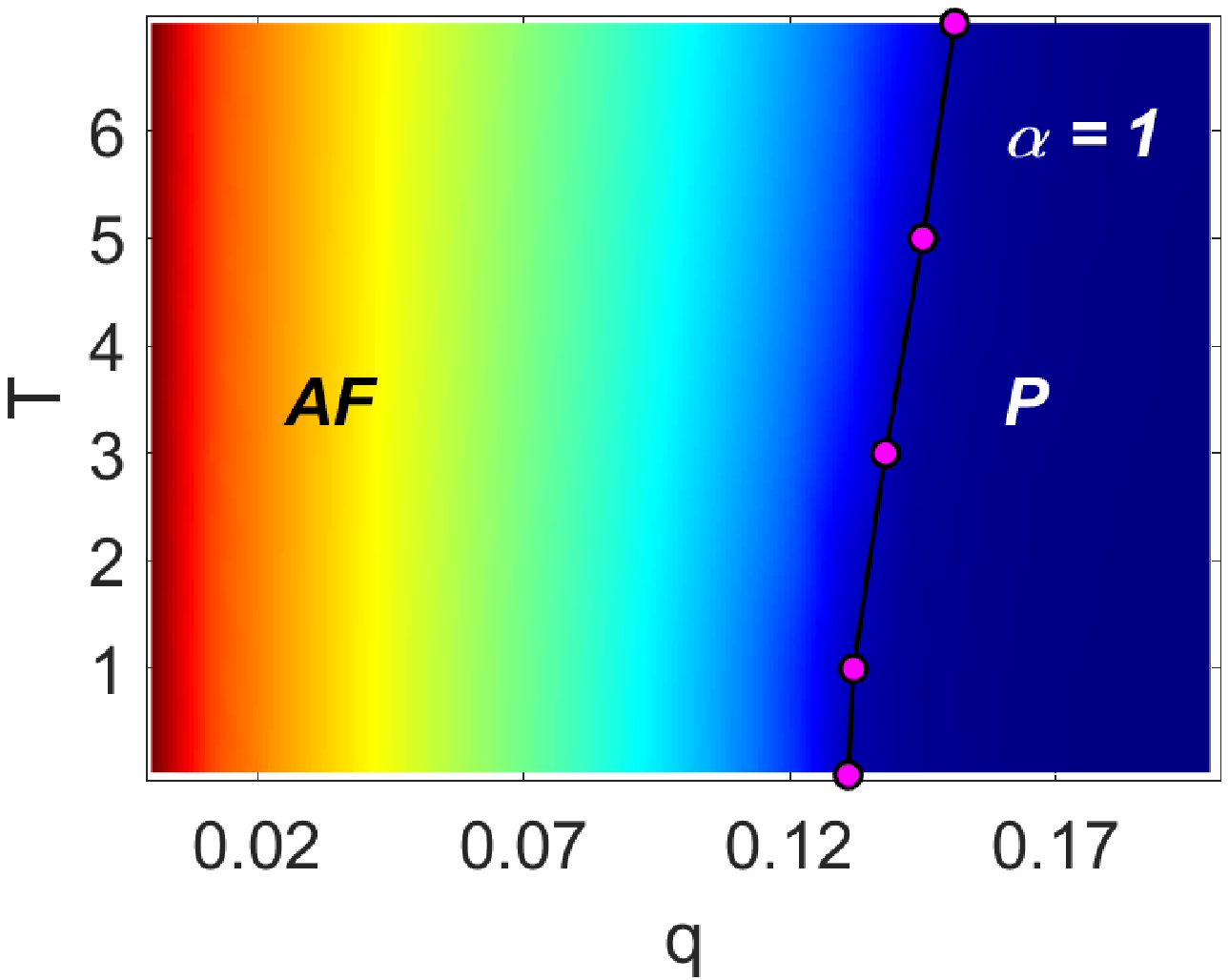}\includegraphics[bb=53bp 0bp 385bp 292bp,clip,scale=0.33]{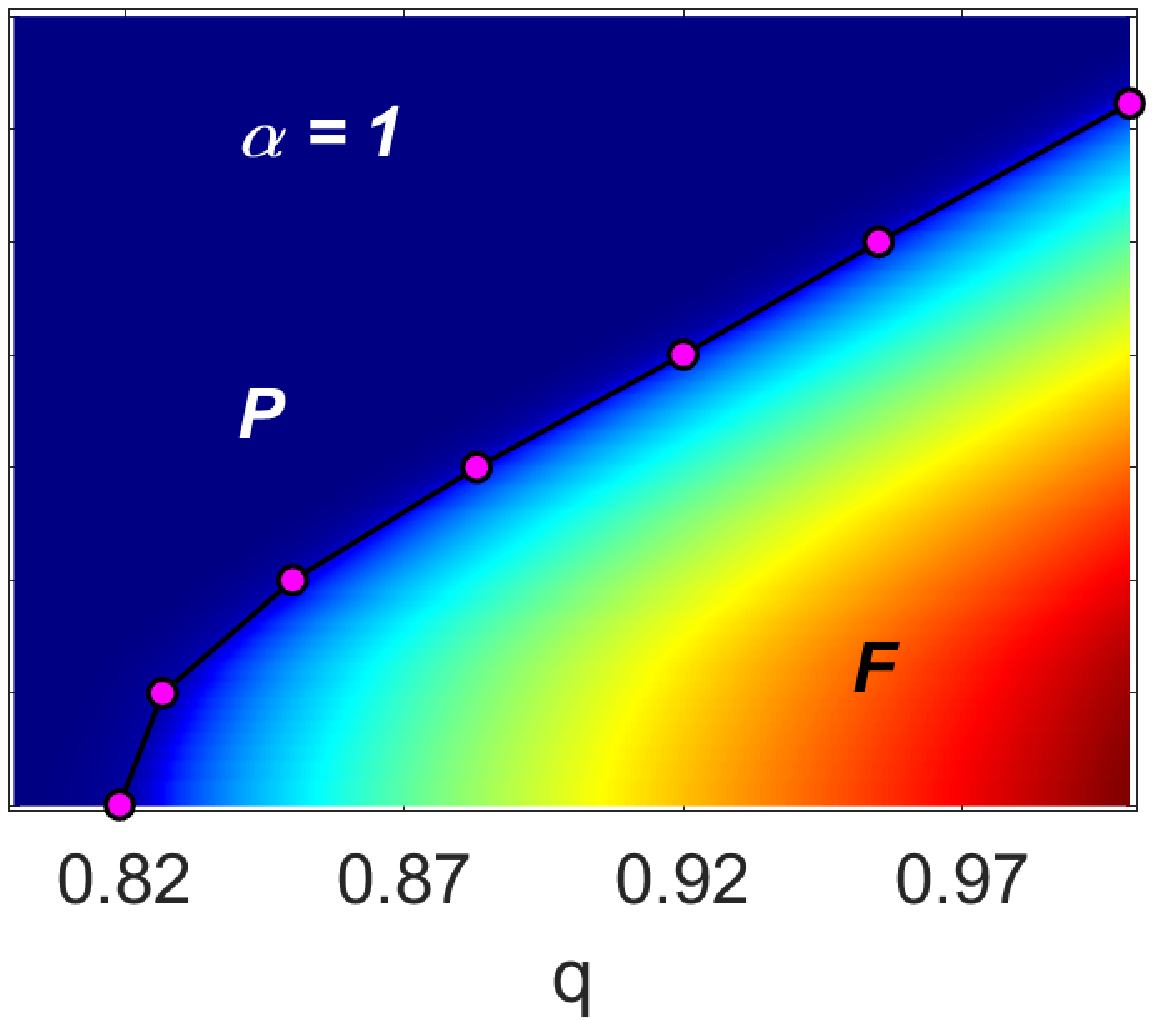}
\par\end{centering}
\caption{{\footnotesize{}Phase diagrams in the plane temperature $T$ versus
$q$ for some values of the exponent $\alpha$ and fixed values of
$k_{0}=4$ and $k_{m}=10$. On the top, we have the color bar of $m_{L}$
where the left side represent the staggered magnetization $\textrm{m}_{\textrm{N}}^{\textrm{AF}}$
illustrating the $AF-P$ transition and the right side the magnetization
per spin $\textrm{m}_{\textrm{N}}^{\textrm{F}}$ illustrating the
$F-P$ transition. The magenta circles are the critical points estimated
by the crossing of $\textrm{U}_{\textrm{L}}$ curves and the black
solid lines are just a guide for the eyes indicating the phase transition
lines. \label{fig:3}}}
\end{figure}

In this section, we present and discuss the results of the nonequilibrium
Ising model on a restricted scale-free network. For the two dynamic
processes, we have an adjustable parameter $q$ that controls the
dynamic competition in the system. If $0<q<1$, the two dynamic processes
have a non-null probability to be chosen and acting in the system,
making it irreversible with respect to the temporal evolution of its
states. As these processes favor the states of higher and lower energy
of the system, with the competition is possible to find stationary
states in the $AF$, $F$, and $P$ phases, based on the Hamiltonian
of the system, Eq. (\ref{eq:1}). With this, it is worth noting that
to obtain a self-organization phenomenon passing from a $F$ to $P$
and from $P$ to $AF$ phases, the division of the network into two
sublattices is essential, once that for nonfrustrated antiparallelism
we must to have well-defined who the central spins are, and to whom
they can connect in the network.

\begin{figure*}
\begin{centering}
\includegraphics[scale=0.6]{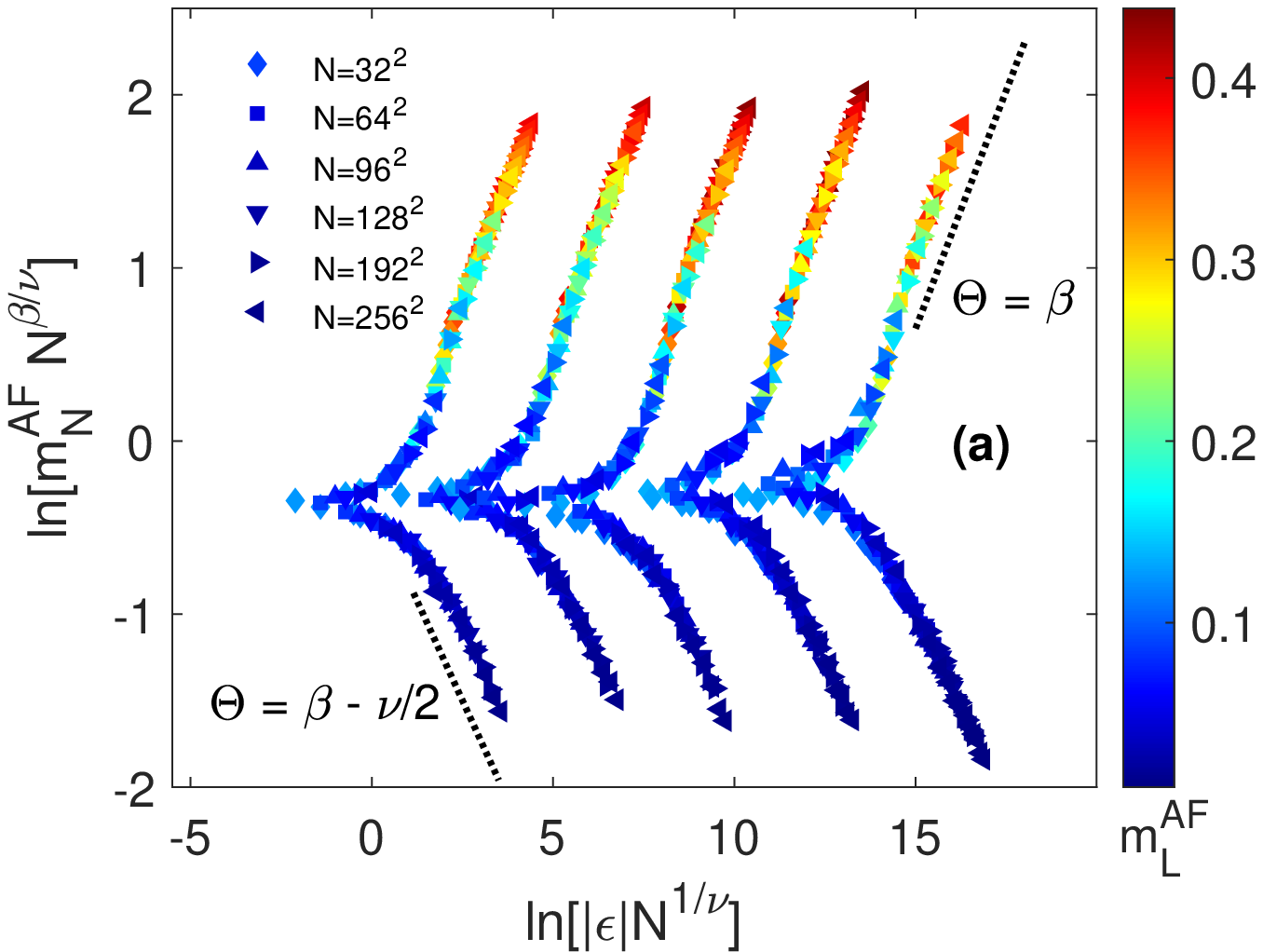}\hspace{0.75cm}\includegraphics[scale=0.6]{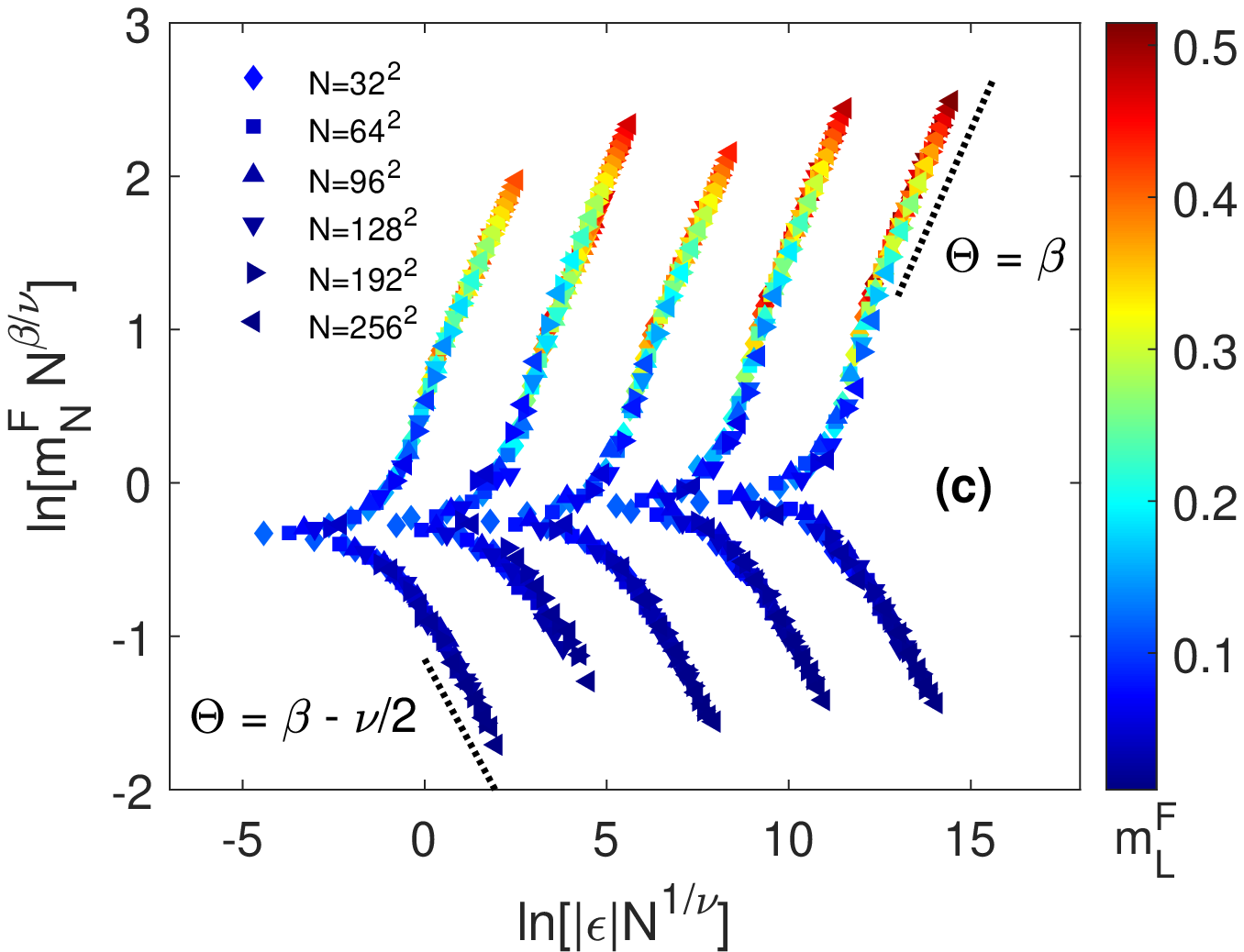}
\par\end{centering}
\begin{centering}
\includegraphics[scale=0.6]{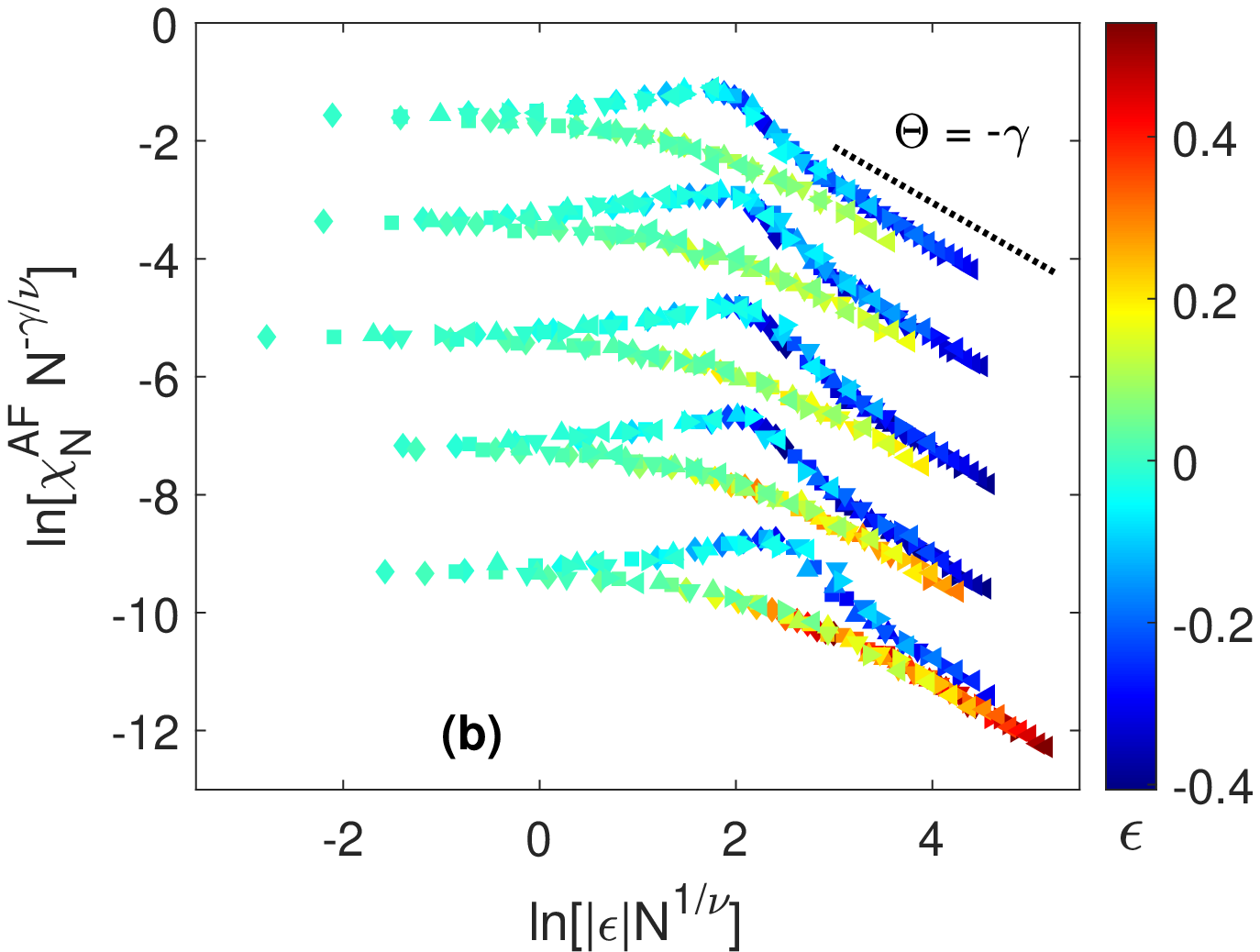}\hspace{0.75cm}\includegraphics[scale=0.6]{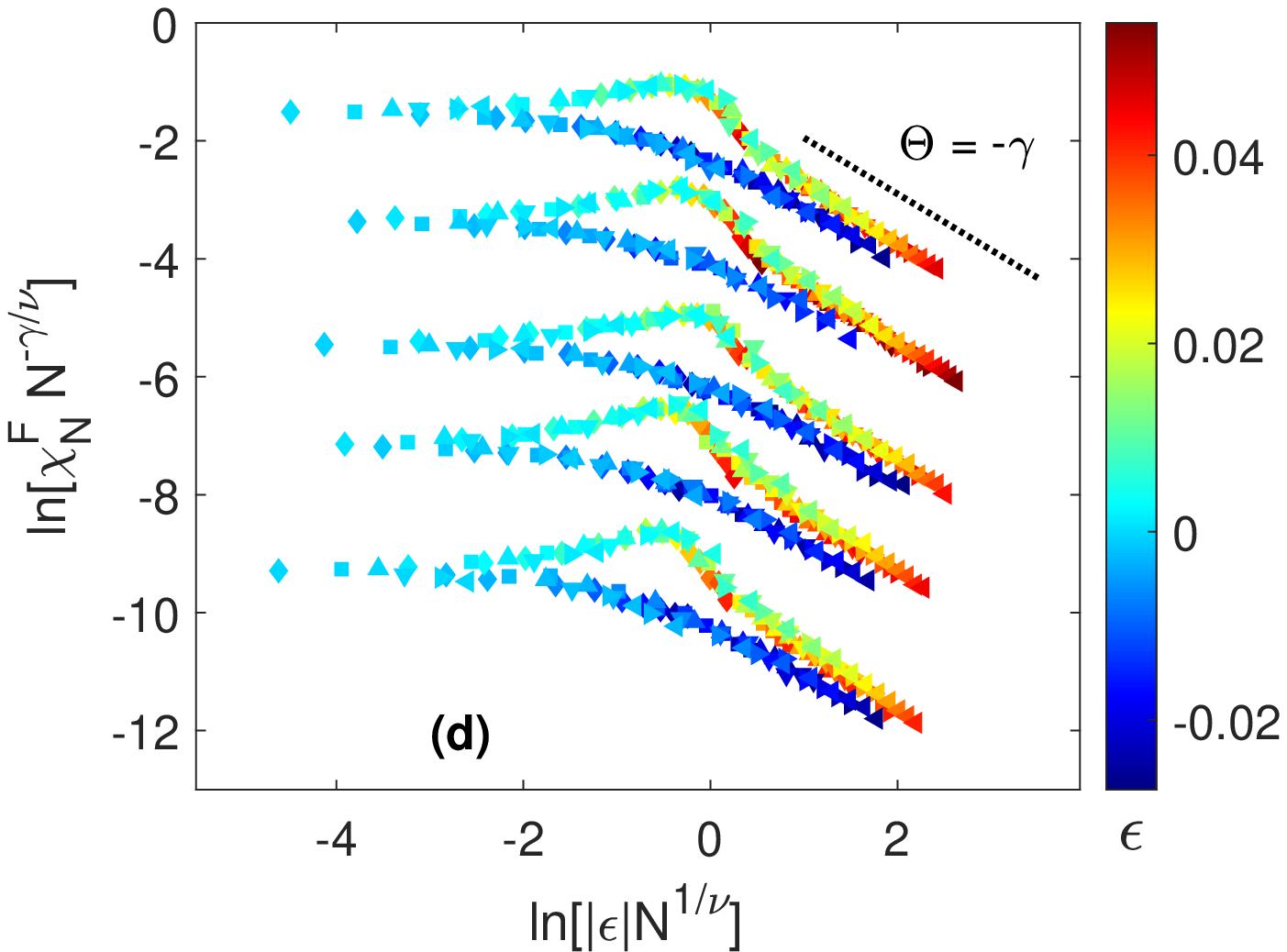}
\par\end{centering}
\caption{{\footnotesize{}Data collapse of $\textrm{m}_{\textrm{L}}^{\textrm{AF}}$
(a), $\textrm{\ensuremath{\chi}}_{\textrm{L}}^{\textrm{AF}}$ (b),
$\textrm{m}_{\textrm{L}}^{\textrm{F}}$ (c) and $\textrm{\ensuremath{\chi}}_{\textrm{L}}^{\textrm{F}}$
(d) for different network sizes as presented in the figures. In (a)
and (c) from the left to the right side we have respectively $\alpha=1,2,3,4$
and $5$, and $\textrm{m}_{\textrm{L}}^{\textrm{AF}}$ and $\textrm{m}_{\textrm{L}}^{\textrm{F}}$color
bars as shown in the figures. In (b) and (d), with $\epsilon=(q-q_{c})/q_{c}$
color bar and from the top to bottom we have the collapses with $\alpha=1,2,3,4$
and $5$, respectively. In these figures, we have changed the positions
curves for $1<\alpha\protect\leq5$ in order to compare all the collapses
obtained. The critical exponents and critical points used here can
be seen in Table \ref{tab:1} and Table \ref{tab:2} respectively.
Here, we have fixed $T=1$, $k_{0}=4$, $k_{m}=10$.\label{fig:4}}}
\end{figure*}

Therefore, the first results can be seen in Fig. \ref{fig:2}, where
we have displayed the thermodynamic quantities obtained with Eqs.
(\ref{eq:7}), (\ref{eq:8}), (\ref{eq:9}) and (\ref{eq:10}). These
quantities were calculated as a function of the competition parameter
$q$, in which is verified that for lower values of $q$ we found
an $AF$ phase, and for higher values of $q$, an $F$ phase is observed.
These phases are easily explained when we look at the dynamics, once
that for lower values of $q$, the two-spin flip mechanism prevails
and this favors the state of high energy in the system, which based
on the ferromagnetic Ising model Hamiltonian is the one where the
spin states are antiparallel, i. e., $AF$ phase. This $AF$ phase
is made explicit in Fig. \ref{fig:2}(a) with the $\textrm{m}_{\textrm{N}}^{\textrm{AF}}$
curves, and with this magnetization is calculated $\textrm{U}_{\textrm{N}}^{\textrm{AF}}$
present in Fig. \ref{fig:2}(b), and its susceptibility $\textrm{\ensuremath{\chi}}_{\textrm{N}}^{\textrm{AF}}$
in Fig. \ref{fig:2}(c). On the other hand, for higher values of $q$,
the one-spin flip mechanism prevails, and as it favors the states
of lower energy in the system, i.e., all spins in the same state,
a ordered phase is also observed, $F$ phase. The quantities related
to this phase is specifically the magnetization $\textrm{m}_{\textrm{N}}^{\textrm{F}}$
curves in Fig. \ref{fig:2}(d), and the $\textrm{U}_{\textrm{N}}^{\textrm{F}}$
and $\textrm{\ensuremath{\chi}}_{\textrm{N}}^{\textrm{F}}$ curves
in Figs. \ref{fig:2}(e) and \ref{fig:2}(f), respectively.

\begin{figure}
\begin{centering}
\includegraphics[scale=0.6]{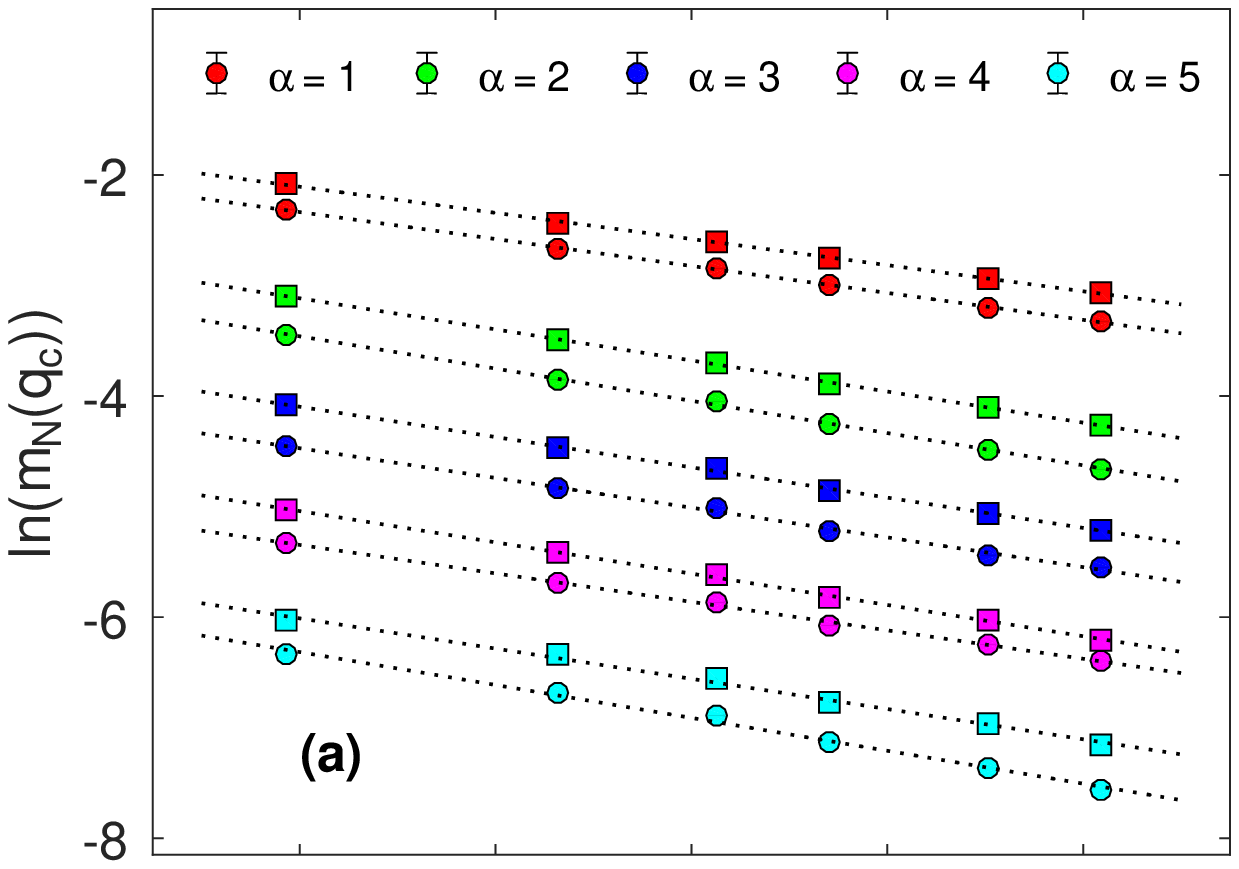}
\par\end{centering}
\begin{centering}
\includegraphics[scale=0.6]{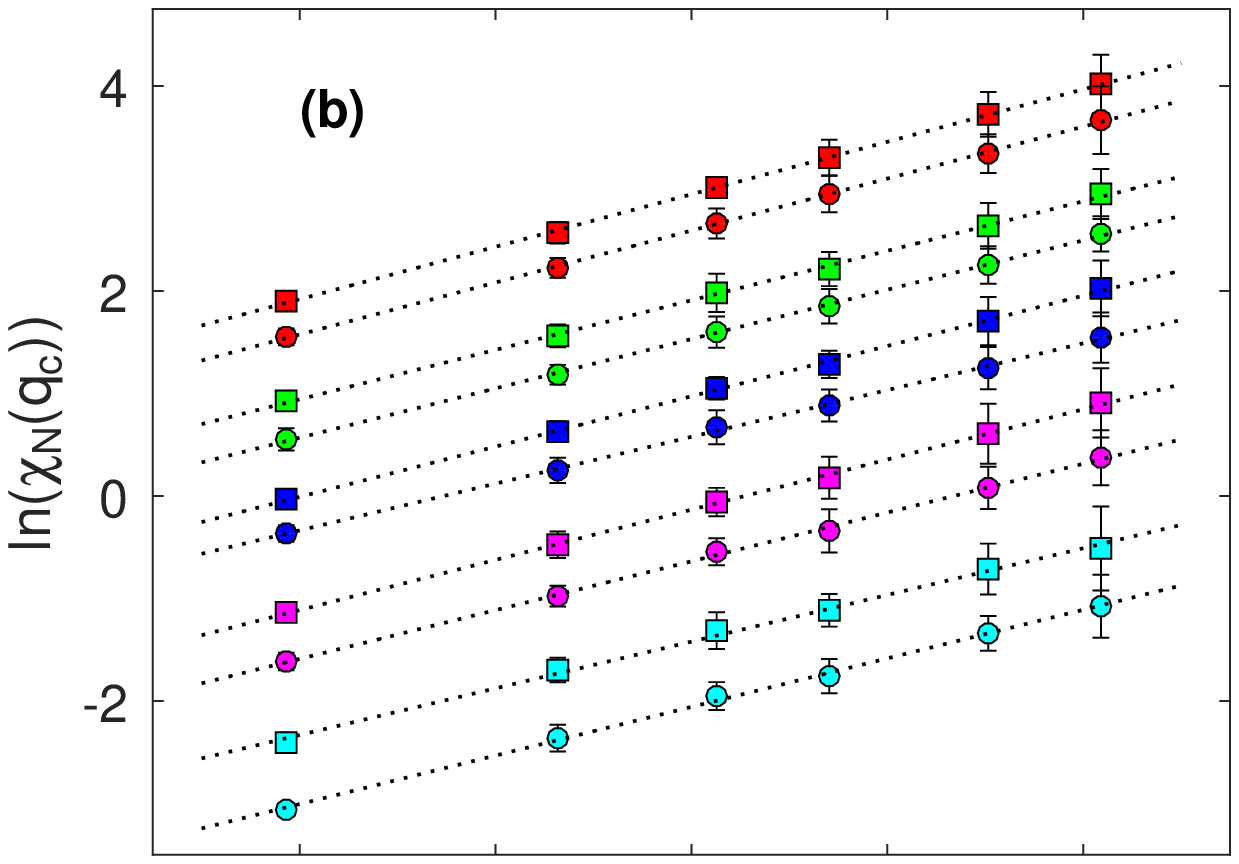}
\par\end{centering}
\begin{centering}
\includegraphics[scale=0.6]{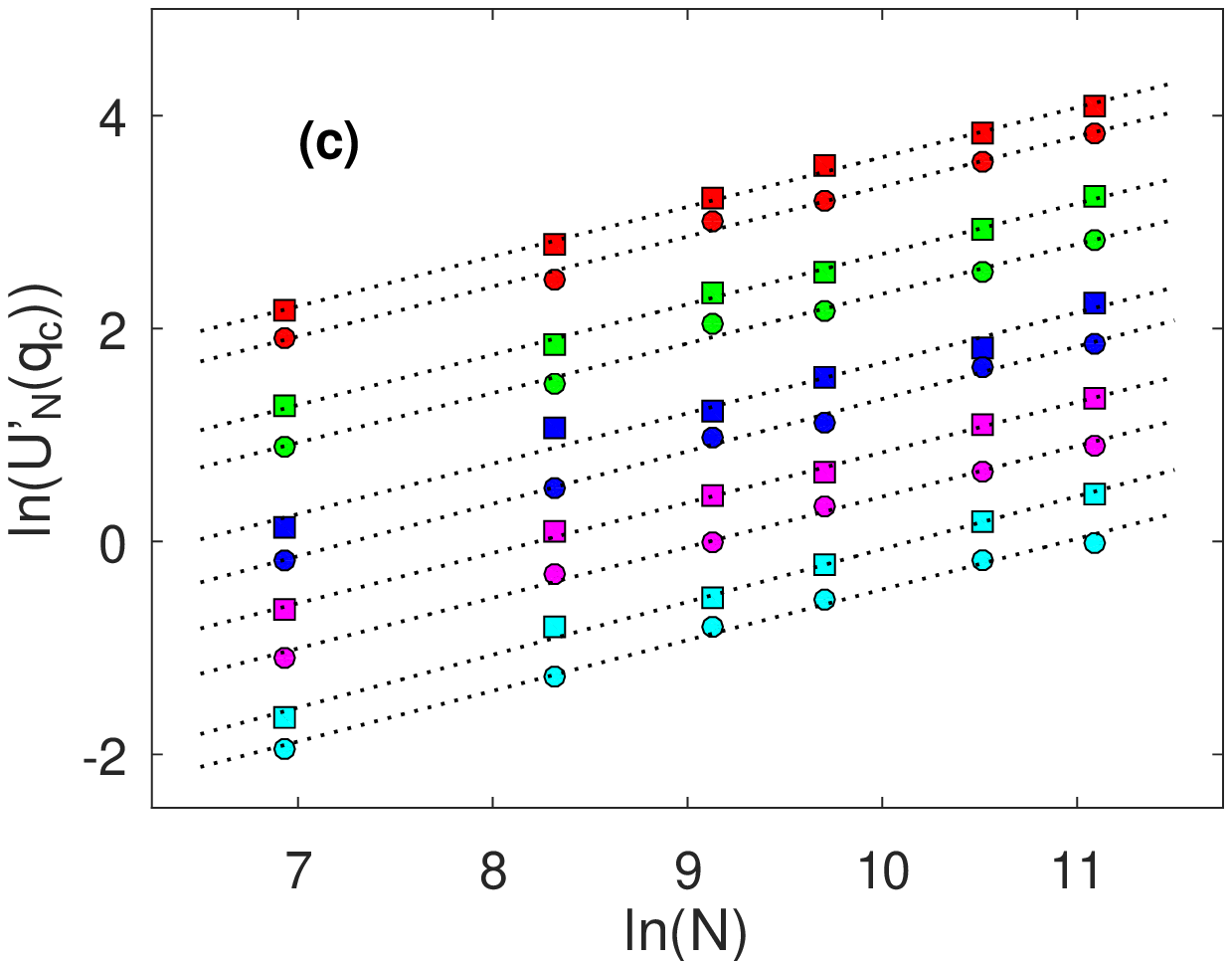}
\par\end{centering}
\caption{{\footnotesize{}In log-log plots is presented linear fits (black dotted
lines) of thermodynamic quantities in the critical point, $\textrm{m}_{\textrm{L}}^{\textrm{AF}}$
$\left(\square\right)$ and $\textrm{m}_{\textrm{L}}^{\textrm{F}}$
$\left(\Circle\right)$ in (a), $\chi_{\textrm{L}}^{\textrm{AF}}$
$\left(\square\right)$ and $\textrm{\ensuremath{\chi}}_{\textrm{L}}^{\textrm{F}}$
$\left(\Circle\right)$ in (b), and $\textrm{U}_{\textrm{L}}^{\textrm{AF}}$
$\left(\square\right)$ and $\textrm{U}_{\textrm{L}}^{\textrm{F}}$
$\left(\Circle\right)$ in (c), both has a function of the network
size $N$, and different values of $\alpha$ as shown in the figures.
Here we have fixed $T=1$, $k_{0}=4$ and $k_{m}=10$. The critical
points used for these fits can be seen in Table \ref{tab:2}. \label{fig:5}}}
\end{figure}

We have used the curves of the fourth-order Binder cumulants for different
network sizes to identify the critical points and order phase transition
\citep{29,30,31,32}. The intersection point of the $\textrm{U}_{\textrm{N}}$
curves indicates the phase transition point on a second-order phase
transition. With the critical point in hand for several values of
adjustable parameters, a phase diagram was built, which can be seen
in Fig. \ref{fig:3}. Therefore, for these diagrams and later results,
we will limit the values $k_{0}=4$ and $k_{m}=10$, once we can build
all networks with sizes ($32)^{2}\le N\le(256)^{2}$, integer exponent
$1\le\alpha\le5$, and compare with others equilibrium \citep{13,14,16}
and nonequilibrium \citep{23,26} Ising model results. Fig. \ref{fig:3}
presents the phase diagrams of temperature $T$ as a function of competition
parameter $q$ for some values of $\alpha$, in which we can see the
$AF$, $F$, and $P$ phases. 

\begin{figure}
\begin{centering}
\includegraphics[scale=0.6]{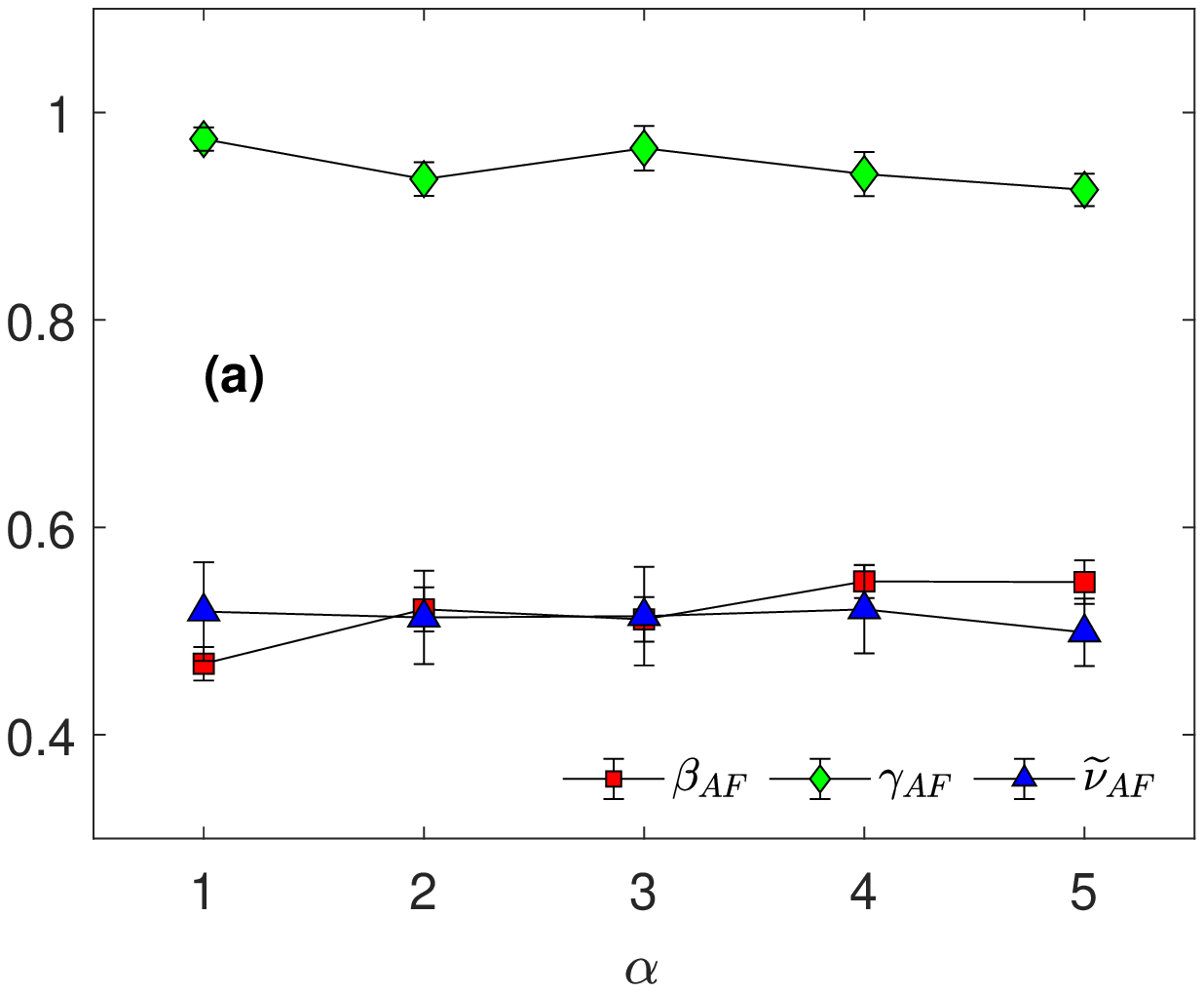}
\par\end{centering}
\begin{centering}
\includegraphics[scale=0.6]{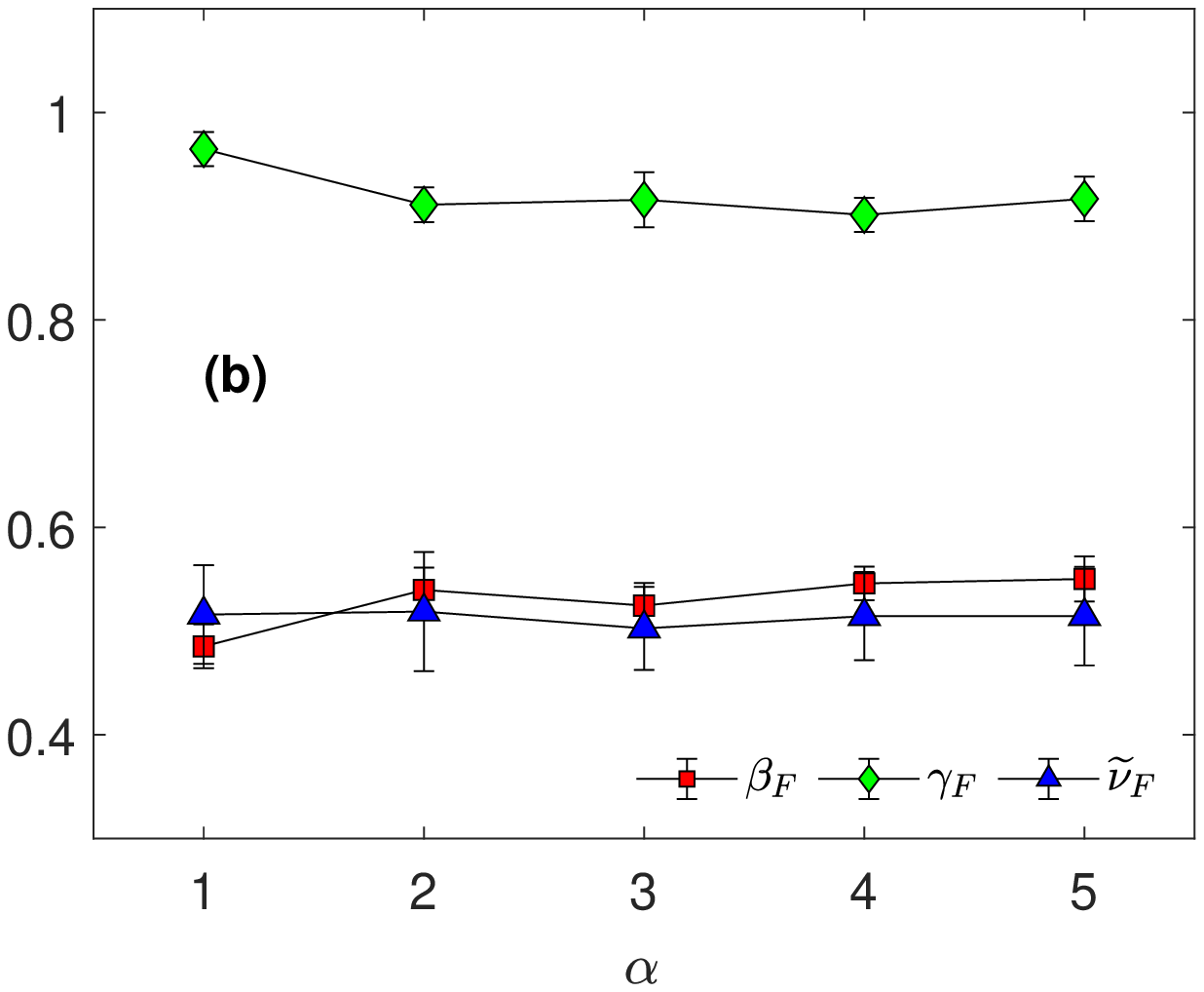}
\par\end{centering}
\caption{{\footnotesize{}Average static critical exponents $\beta$, $\gamma$,
and $\nu$, obtained form the slope of scaling functions and the data
collapse of magnetization and susceptibility curves as a function
of $\alpha$. (a) For $AF-P$ transition and (b) for the $F-p$ transition.
These exponents were obtained with fixed values of $T=1$, $k_{0}=4$
and $k_{m}=10$. \label{fig:6}}}
\end{figure}

In these diagrams (Fig. \ref{fig:3}), we have illustrated the self-organization
phenomena with the transitions between $AF$ to $P$ phases, and $P$
to $F$ phases. Since the scale is fixed in all the figures, we can
also see that when we decrease the value of $\alpha$, the region
of ordered phases, $F$ and $AF$, increases. This change in the topology
of the diagram is related to the degree distribution, once the lower
values of exponent $\alpha$ mean a high probability of having more
connected sites on network, i.e., more sites with a degree $k_{m}$.
Consequently, knowing that more connected sites on the stationary
ordered state require more energy to override its interactions, larger
are the regions of the ordered phases. Another interesting observation
that we can do, is regarding the shape of the regions in the ordered
phases. The ferromagnetic phases are driven by the one-spin flip mechanism
described by Metropolis prescription, which is very dependent on $T$,
and for high $T$ we observe the disordered phase $P$. On the other
hand, $AF$ phases is driven by the two-spin flip mechanism, in which
is a simpler process and little influenced by temperature.

All systems belonging to a given universality class share the same
set of critical exponents. The critical points can be used to describe
the critical behavior in the sense of universality class with the
set of critical exponents. Here, we have computed the exponents $\beta$,
$\gamma$ and $\nu$, by two methods. The first one is based on the
data collapse, in which we use the scaling relations, Eqs. (\ref{eq:11})
and (\ref{eq:12}), to obtain the scaling functions of magnetization
and susceptibility with its collapsed curves. This is possible because
in the proximity of the critical points the scaling relations are
independent of network size with the correct critical exponents and
critical point of the system \citep{29,30,31}. To obtain the critical
exponents by this method and using the already estimated critical
points, we have plotted the scaling functions $m_{0}(N^{1/\nu}\epsilon)$
and $\chi_{0}(N^{1/\nu}\epsilon)$ as a function of $|\epsilon|N^{1/\nu}$
for different network sizes and in the proximity of critical points.
Therefore, for $\epsilon\to0$ and adjusting the involved critical
exponents, when the curves of different network sizes collapse better
into a single curve, these exponents used are considered the critical
exponents of the system. 

\begin{table*}
\begin{centering}
\begin{tabular}{|c|c|c|c|c|c|c|c|c|}
\hline 
{\footnotesize{}$\alpha$} & {\footnotesize{}$\beta_{F}$} & {\footnotesize{}$\gamma_{F}$} & {\footnotesize{}$\nu_{F}\left(\textrm{m}_{\textrm{L}}\right)$} & {\footnotesize{}$\nu_{F}\left(\textrm{\ensuremath{\chi}}_{\textrm{L}}\right)$} & {\footnotesize{}$\beta_{AF}$} & {\footnotesize{}$\gamma_{AF}$} & {\footnotesize{}$\nu_{AF}\left(\textrm{m}_{\textrm{L}}\right)$} & {\footnotesize{}$\nu_{AF}\left(\textrm{\ensuremath{\chi}}_{\textrm{L}}\right)$}\tabularnewline
\hline 
\hline 
{\footnotesize{}$1$} & $0.51\pm0.04$ & $0.98\pm0.03$ & $1.98\pm0.03$ & $2.02\pm0.04$ & $0.50\pm0.03$ & $1.00\pm0.02$ & $2.00\pm0.04$ & $2.00\pm0.03$\tabularnewline
\hline 
{\footnotesize{}$2$} & $0.54\pm0.04$ & $0.92\pm0.03$ & $2.00\pm0.05$ & $2.00\pm0.04$ & $0.51\pm0.04$ & $0.95\pm0.03$ & $2.00\pm0.05$ & $2.00\pm0.03$\tabularnewline
\hline 
{\footnotesize{}$3$} & $0.52\pm0.04$ & $0.93\pm0.05$ & $2.00\pm0.02$ & $1.96\pm0.02$ & $0.50\pm0.04$ & $1.00\pm0.04$ & $2.00\pm0.03$ & $2.02\pm0.04$\tabularnewline
\hline 
{\footnotesize{}$4$} & $0.56\pm0.03$ & $0.90\pm0.03$ & $1.96\pm0.03$ & $2.06\pm0.04$ & $0.56\pm0.03$ & $0.96\pm0.04$ & $2.02\pm0.04$ & $2.04\pm0.03$\tabularnewline
\hline 
{\footnotesize{}$5$} & $0.54\pm0.04$ & $0.95\pm0.04$ & $1.96\pm0.04$ & $2.06\pm0.03$ & $0.55\pm0.04$ & $0.95\pm0.03$ & $2.01\pm0.02$ & $1.92\pm0.03$\tabularnewline
\hline 
\end{tabular}
\par\end{centering}
\caption{{\footnotesize{}Critical exponents obtained by the data collapse method
for several values $\alpha$. The $F-P$ transitions are denoted by
$F$ subscript and the $AF-P$ transitions are denoted by $AF$ subscript.
In these transitions, the data collapse of magnetization and susceptibility
curves returns us respectively $\nu(\textrm{m}_{\textrm{L}})$ and
$\nu(\textrm{\ensuremath{\chi}}_{\textrm{L}})$ estimates using the
$\nu$ exponent. For these exponents, we have fixed $T=1$ and $k_{0}=4$
and $k_{m}=10$. The collapsed curves can be seen in Fig. \ref{fig:4}.
\label{tab:1}}}
\end{table*}

Fig. \ref{fig:4} display the scaling functions $m_{0}(N^{1/\nu}\epsilon)$
and $\chi_{0}(N^{1/\nu}\epsilon)$ collapsed in the log-log plot to
obtain its asymptotic behavior. In these figures, we have fixed $T=1$
to obtain the critical exponents of the system both in the $F-P$
transition and in the $AF-P$ transition, for all values of $\alpha$.
In Fig. \ref{fig:4}(a), we can see the function $m_{0}(N^{1/\nu}\epsilon)$
in the log-log plot, produced with the collapse of $\textrm{m}_{\textrm{N}}^{\textrm{AF}}$
curves, and with this was obtained the exponents $\beta$ and $\nu$.
In the same way, in Fig. \ref{fig:4}(b) the scaling function $\chi_{0}(N^{1/\nu}\epsilon)$
is presented in the log-log plot with $\textrm{\ensuremath{\chi}}_{\textrm{N}}$
curves based on staggered magnetization, in which with the best data
collapse we have obtained the exponent $\gamma$, and another estimated
value for the $\nu$ exponent. On the other hand, in the $F-P$ transition,
Figs. \ref{fig:4}(c) and \ref{fig:4}(d), respectively, contain the
log-log plot of the scaling functions based on $\textrm{m}_{\textrm{N}}^{\textrm{F}}$
and its susceptibility, $\textrm{\ensuremath{\chi}}_{\textrm{N}}^{\textrm{F}}$.
The asymptotic behavior, away from the critical point of these functions,
is predicted to a slope $\Theta$ related to the obtained critical
exponents, once that for the magnetization curves starting from the
ordered phase, below from the critical point $\Theta=\beta$, and
above it $\Theta=\nu/2-\beta$, and for the susceptibility curves
we only have $\Theta=-\gamma$. The critical exponents obtained by
this first method are presented in Table \ref{tab:1} and the critical
point used for them can be seen in Table \ref{tab:2}.

\begin{table*}
\begin{centering}
\begin{tabular}{|c|c|c|c|c|c|c|c|c|}
\hline 
{\footnotesize{}$\alpha$} & {\footnotesize{}$q_{c}^{F}$} & {\footnotesize{}$q_{c}^{AF}$} & {\footnotesize{}$\left(-\beta/\nu\right)_{F}$} & {\footnotesize{}$\left(\gamma/\nu\right)_{F}$} & {\footnotesize{}$\left(1/\nu\right)_{F}$} & {\footnotesize{}$\left(-\beta/\nu\right)_{AF}$} & {\footnotesize{}$\left(\gamma/\nu\right)_{AF}$} & {\footnotesize{}$\left(1/\nu\right)_{AF}$}\tabularnewline
\hline 
\hline 
{\footnotesize{}$1$} & $0.8267\pm0.0002$ & $0.1320\pm0.0002$ & $0.24\pm0.04$ & $0.51\pm0.05$ & $0.47\pm0.06$ & $0.24\pm0.04$ & $0.51\pm0.04$ & $0.47\pm0.06$\tabularnewline
\hline 
{\footnotesize{}$2$} & $0.8396\pm0.0003$ & $0.1156\pm0.0002$ & $0.28\pm0.04$ & $0.48\pm0.05$ & $0.47\pm0.07$ & $0.29\pm0.05$ & $0.48\pm0.05$ & $0.47\pm0.05$\tabularnewline
\hline 
{\footnotesize{}$3$} & $0.8534\pm0.0004$ & $0.0995\pm0.0005$ & $0.27\pm0.06$ & $0.49\pm0.05$ & $0.47\pm0.06$ & $0.27\pm0.05$ & $0.46\pm0.05$ & $0.49\pm0.06$\tabularnewline
\hline 
{\footnotesize{}$4$} & $0.8664\pm0.0004$ & $0.0842\pm0.0004$ & $0.28\pm0.05$ & $0.49\pm0.06$ & $0.47\pm0.05$ & $0.26\pm0.04$ & $0.48\pm0.05$ & $0.48\pm0.05$\tabularnewline
\hline 
{\footnotesize{}$5$} & $0.8788\pm0.0004$ & $0.0714\pm0.0004$ & $0.27\pm0.06$ & $0.46\pm0.05$ & $0.50\pm0.06$ & $0.30\pm0.05$ & $0.47\pm0.03$ & $0.48\pm0.04$\tabularnewline
\hline 
\end{tabular}
\par\end{centering}
\caption{{\footnotesize{}Critical points obtained by the intersection of $\textrm{U}_{\textrm{L}}$
curves, and the critical exponents by the second method for several
values $\alpha$. The $F-P$ transitions are denoted by $F$ subscript
and the $AF-P$ transitions are denoted by $AF$ subscript. The slope
of the straight lines are presented in Fig, \ref{fig:5}. \label{tab:2}}}
\end{table*}

Now, let use a second method to calculate the critical exponents and
also using the scaling relations, but, with the log-log plot of $\textrm{m}_{\textrm{N}}^{\textrm{AF}}$
and $\textrm{m}_{\textrm{N}}^{\textrm{F}}$ at its respective $\textrm{\ensuremath{\chi}}_{\textrm{N}}$
and $\textrm{U}_{\textrm{N}}$ in the proximity of the critical point
as a function of $N$. The slope on this set of point returns us specific
ratios between the critical exponents. Fig. \ref{fig:5}(a) shown
the points of $\textrm{m}_{\textrm{N}}^{\textrm{AF}}$ and $\textrm{m}_{\textrm{N}}^{\textrm{F}}$
in the vicinity of the critical point as a function of network sizes
$N$. With the best fit of these points and its slope based on the
scaling relation of Eq. (\ref{eq:11}), gives us the estimate of the
ratio $-\beta/\nu$. In the same way, but for the susceptibility of
these magnetizations, on the vicinity of the critical point as a function
of $N$ in the log-log plot, is presented in Fig. \ref{fig:5}(b).
The best fit with the points in this figure gives us the slope related
to the ratio $\gamma/\nu$ presented in the scaling relation of Eq.
(\ref{eq:12}). The ratio between these critical exponents is interesting
but does not reveal the correct value of the exponents separately.
Thus, to solve this, we used the scaling relation in Eq. (\ref{eq:13}),
in which the derivative of $\textrm{U}_{\textrm{N}}$ in the vicinity
of the critical point and different network sizes gives us the ratio
$1/\nu$. This ratio is illustrated in Fig. \ref{fig:5}(c) by its
log-log plot. All the ratio between the critical exponents obtained
on this method can be found in Table \ref{tab:2}.

From these two used methods are obtained equivalent exponents. But,
we have to pay attention that as we are dealing with random interactions
on the network, we do not have a well-defined dimension, and consequently,
it was necessary to use scaling relations dependent only of the system
size, Eqs. (\ref{eq:11}), (\ref{eq:12}) and (\ref{eq:13}). Therefore,
the expected mean-field finite-size scaling exponents due to these
equations are $\beta=1/2$, $\gamma=1$, and $\nu=2$ \citep{28}.
If compared to the usual Ising model mean-field exponents $\beta=\widetilde{\nu}=1/2$,
and $\gamma=1$, the only exponent affected by the dimension of the
system is the related to the correlation length, $\nu$, but, can
be derived by the relation $\nu=d_{u}\widetilde{\nu}$, where $d_{u}$
is the upper critical dimension, that in the Ising model is $d_{u}=4$.
With these information, we have computed $\widetilde{\nu}$ based
on the exponents $\nu$ obtained here. The critical exponents of the
system, obtained by the two methods are compiled in Fig. \ref{fig:6}(a)
for $AF-P$ transitions, and in Fig. \ref{fig:6}(b) for $F-P$  transitions,
both as a function of $\alpha$. Comparing these obtained critical
exponents with the mean-field critical exponents, we can see that
for $\alpha=1$ is obtained the more accurate mean-field critical
exponents, however, as $\alpha$ increase, the critical exponents
are still of mean-field but with a little deviation. This deviation
was explained in the work with the equilibrium Ising model on a restricted
scale-free network \citep{16}, and is due to the increase of degree-degree
correlations \citep{33} with the decreasing of more connected sites.

For the sake of curiosity, our network was labeled as a square lattice,
which we always use $N=L^{2}$ sites. Thus, changing $N$ in Eqs.
(\ref{eq:11}), (\ref{eq:12}) and (\ref{eq:13}) by $L$, we have
the scaling relations depending on the dimension of the system, that
in our case is two dimensions. With these new scaling relations, we
have computed again the critical exponents of the system and we have
obtained the same critical exponents of systems upper the Ising model
critical dimension, by adding long-range interactions on a regular
square lattice \citep{15,26}. It indicates that with our selected
network sizes $N=L^{2}$, we also could use the scaling relations
depending on the dimension of the system to calculate the critical
exponents. However, when dealing with complex networks, this dimensioning
possibility is not always available, once that the objective is to
model real networks \citep{12,34,35}. In this case, Hong \emph{et
al.} \citep{28} proposed scaling relations for complex networks independent
of system dimension, and from them, obtained the set of mean-field
finite-size-scaling exponents.

\section{Conclusions\label{sec:Conclusions}}

Here, we have employed Monte Carlo simulations to study the thermodynamic
quantities and the critical behavior of the nonequilibrium Ising model
on a restricted scale-free network. By using one- and two-spin flip
competing dynamics, we reach the stationary state of the system at
the nonequilibrium regime. Fixing the maximum and minimum degree values
for the whole network size and by using FSS analysis, we are able
to always find a finite critical point even being in a network with
power-law degree distribution, since we always have second and fourth
convergent moments based on its distribution $P(k)$. As a result,
we have obtained the critical points from the second-order phase transitions
based on the intersection of $\textrm{U}_{\textrm{N}}$ curves and
built a phase diagram of temperature $T$ as a function of the competition
parameter $q$. In these diagrams, we have verified the self-organization
phenomena in the transitions from $AF$ to $P$ phases in lower values
of $q$, and from $P$ to $F$ phases in higher values of $q$ and
lower $T$. Because we are dealing with a power-law degree distribution
on the network, $P(k)\sim k^{-\alpha}$, decreasing the value of $\alpha$,
increase the number of more connected sites, and as consequence, also
increase the region of the ordered phases in the diagram. Topologies
equivalent to these diagrams were also obtained in previous works
with the same dynamics, but in different networks and models \citep{23,26}.
Through FSS arguments, we calculated the critical exponents $\beta$,
$\gamma$, and $\nu$ for the system, and as a function of $\alpha$,
because we have a restricted scale-free network in which its second
and fourth moments of degree distribution are convergent. In this
case, we have always found the mean-field critical exponents and a
slight deviation from them with the increasing degree-degree correlations.
This mean-field behavior follows the predicted and observed critical
behavior in other complex networks \citep{13,14,15,16,17}, in addition
to being another agreement of what was conjectured by Grinstein \emph{et
al}. \citep{24}, i.e., we obtained the same universality class of
the Ising model on a restricted scale-free network both in the equilibrium
regime \citep{16} as out of it.

\end{document}